\documentclass[reprint,
superscriptaddress,
 amsmath,
 amssymb,
 aps,
preprintnumbers,
]{revtex4-1}

\usepackage{aasmacros}
\usepackage{graphicx}
\usepackage{dcolumn}
\usepackage{bm}
\usepackage{mathrsfs}
\usepackage{xfrac}
\usepackage{enumitem}
\usepackage[mathlines]{lineno}

\usepackage{color}

\def\defDIA{$D_{\mathrm{IA}} = 1 - 0.11(A_{\mathrm{IA}}-1)$}

\newcommand{\norm}[1]{\left\lVert#1\right\rVert}

\def\band{$i$}
\def\nsurveys{30}
\def\e#1{{\cdot 10^{#1}}}

\usepackage{multirow}
\usepackage{tabularx}
\usepackage{rotating}

\begin{document}

\title{Monte Carlo Control Loops for cosmic shear cosmology with DES Year 1}

\author{T.~Kacprzak}
\email[Corresponding author: ]{tomasz.kacprzak@phys.ethz.ch}
\affiliation{Institute for Particle Physics and Astrophysics, ETH Zurich, Wolfgang-Pauli-Strasse 27, CH-8093 Zurich, Switzerland}
\author{J.~Herbel}
\affiliation{Institute for Particle Physics and Astrophysics, ETH Zurich, Wolfgang-Pauli-Strasse 27, CH-8093 Zurich, Switzerland}
\author{A.~Nicola}
\affiliation{Department of Astrophysical Sciences, Princeton University, Princeton, NJ 08544, USA}
\author{R.~Sgier}
\affiliation{Institute for Particle Physics and Astrophysics, ETH Zurich, Wolfgang-Pauli-Strasse 27, CH-8093 Zurich, Switzerland}
\author{F.~Tarsitano}
\affiliation{Institute for Particle Physics and Astrophysics, ETH Zurich, Wolfgang-Pauli-Strasse 27, CH-8093 Zurich, Switzerland}
\author{C.~Bruderer}
\affiliation{Institute for Particle Physics and Astrophysics, ETH Zurich, Wolfgang-Pauli-Strasse 27, CH-8093 Zurich, Switzerland}
\author{A.~Amara}
\affiliation{Institute for Particle Physics and Astrophysics, ETH Zurich, Wolfgang-Pauli-Strasse 27, CH-8093 Zurich, Switzerland}
\author{A.~Refregier}
\affiliation{Institute for Particle Physics and Astrophysics, ETH Zurich, Wolfgang-Pauli-Strasse 27, CH-8093 Zurich, Switzerland}
\author{S.~L.~Bridle}
\affiliation{Jodrell Bank Center for Astrophysics, School of Physics and Astronomy, University of Manchester, Oxford Road, Manchester, M13 9PL, UK}
\author{A.~Drlica-Wagner}
\affiliation{Fermi National Accelerator Laboratory, P. O. Box 500, Batavia, IL 60510, USA}
\affiliation{Kavli Institute for Cosmological Physics, University of Chicago, Chicago, IL 60637, USA}
\author{D.~Gruen}
\affiliation{Department of Physics, Stanford University, 382 Via Pueblo Mall, Stanford, CA 94305, USA}
\affiliation{Kavli Institute for Particle Astrophysics \& Cosmology, P. O. Box 2450, Stanford University, Stanford, CA 94305, USA}
\affiliation{National Accelerator Laboratory, Menlo Park, CA 94025, USA}
\author{W.~G.~Hartley}
\affiliation{Department of Physics \& Astronomy, University College London, Gower Street, London, WC1E 6BT, UK}
\author{B.~Hoyle}
\affiliation{Max Planck Institute for Extraterrestrial Physics, Giessenbachstrasse, 85748 Garching, Germany}
\affiliation{Universit\"ats-Sternwarte, Fakult\"at f\"ur Physik, Ludwig-Maximilians Universit\"at M\"unchen, Scheinerstr. 1, 81679 M\"unchen, Germany}
\author{L.~F.~Secco}
\affiliation{Department of Physics and Astronomy, University of Pennsylvania, Philadelphia, PA 19104, USA}
\author{J.~Zuntz}
\affiliation{Institute for Astronomy, University of Edinburgh, Edinburgh EH9 3HJ, UK}
\author{J.~Annis}
\affiliation{Fermi National Accelerator Laboratory, P. O. Box 500, Batavia, IL 60510, USA}
\author{S.~Avila}
\affiliation{Instituto de Fisica Teorica UAM/CSIC, Universidad Autonoma de Madrid, 28049 Madrid, Spain}
\author{E.~Bertin}
\affiliation{CNRS, UMR 7095, Institut d'Astrophysique de Paris, F-75014, Paris, France}
\affiliation{Sorbonne Universit\'es, UPMC Univ Paris 06, UMR 7095, Institut d'Astrophysique de Paris, F-75014, Paris, France}
\author{D.~Brooks}
\affiliation{Department of Physics \& Astronomy, University College London, Gower Street, London, WC1E 6BT, UK}
\author{E.~Buckley-Geer}
\affiliation{Fermi National Accelerator Laboratory, P. O. Box 500, Batavia, IL 60510, USA}
\author{A.~Carnero~Rosell}
\affiliation{Centro de Investigaciones Energ\'eticas, Medioambientales y Tecnol\'ogicas (CIEMAT), Madrid, Spain}
\affiliation{Laborat\'orio Interinstitucional de e-Astronomia - LIneA, Rua Gal. Jos\'e Cristino 77, Rio de Janeiro, RJ - 20921-400, Brazil}
\author{M.~Carrasco~Kind}
\affiliation{Department of Astronomy, University of Illinois at Urbana-Champaign, 1002 W. Green Street, Urbana, IL 61801, USA}
\affiliation{National Center for Supercomputing Applications, 1205 West Clark St., Urbana, IL 61801, USA}
\author{J.~Carretero}
\affiliation{Institut de F\'{\i}sica d'Altes Energies (IFAE), The Barcelona Institute of Science and Technology, Campus UAB, 08193 Bellaterra (Barcelona) Spain}
\author{L.~N.~da Costa}
\affiliation{Laborat\'orio Interinstitucional de e-Astronomia - LIneA, Rua Gal. Jos\'e Cristino 77, Rio de Janeiro, RJ - 20921-400, Brazil}
\affiliation{Observat\'orio Nacional, Rua Gal. Jos\'e Cristino 77, Rio de Janeiro, RJ - 20921-400, Brazil}
\author{J.~De~Vicente}
\affiliation{Centro de Investigaciones Energ\'eticas, Medioambientales y Tecnol\'ogicas (CIEMAT), Madrid, Spain}
\author{S.~Desai}
\affiliation{Department of Physics, IIT Hyderabad, Kandi, Telangana 502285, India}
\author{H.~T.~Diehl}
\affiliation{Fermi National Accelerator Laboratory, P. O. Box 500, Batavia, IL 60510, USA}
\author{P.~Doel}
\affiliation{Department of Physics \& Astronomy, University College London, Gower Street, London, WC1E 6BT, UK}
\author{J.~Garc\'ia-Bellido}
\affiliation{Instituto de Fisica Teorica UAM/CSIC, Universidad Autonoma de Madrid, 28049 Madrid, Spain}
\author{E.~Gaztanaga}
\affiliation{Institut d'Estudis Espacials de Catalunya (IEEC), 08034 Barcelona, Spain}
\affiliation{Institute of Space Sciences (ICE, CSIC),  Campus UAB, Carrer de Can Magrans, s/n,  08193 Barcelona, Spain}
\author{R.~A.~Gruendl}
\affiliation{Department of Astronomy, University of Illinois at Urbana-Champaign, 1002 W. Green Street, Urbana, IL 61801, USA}
\affiliation{National Center for Supercomputing Applications, 1205 West Clark St., Urbana, IL 61801, USA}
\author{J.~Gschwend}
\affiliation{Laborat\'orio Interinstitucional de e-Astronomia - LIneA, Rua Gal. Jos\'e Cristino 77, Rio de Janeiro, RJ - 20921-400, Brazil}
\affiliation{Observat\'orio Nacional, Rua Gal. Jos\'e Cristino 77, Rio de Janeiro, RJ - 20921-400, Brazil}
\author{G.~Gutierrez}
\affiliation{Fermi National Accelerator Laboratory, P. O. Box 500, Batavia, IL 60510, USA}
\author{D.~L.~Hollowood}
\affiliation{Santa Cruz Institute for Particle Physics, Santa Cruz, CA 95064, USA}
\author{K.~Honscheid}
\affiliation{Center for Cosmology and Astro-Particle Physics, The Ohio State University, Columbus, OH 43210, USA}
\affiliation{Department of Physics, The Ohio State University, Columbus, OH 43210, USA}
\author{D.~J.~James}
\affiliation{Center for Astrophysics $\vert$ Harvard \& Smithsonian, 60 Garden Street, Cambridge, MA 02138, USA}
\author{M.~Jarvis}
\affiliation{Department of Physics and Astronomy, University of Pennsylvania, Philadelphia, PA 19104, USA}
\author{M.~Lima}
\affiliation{Departamento de F\'isica Matem\'atica, Instituto de F\'isica, Universidade de S\~ao Paulo, CP 66318, S\~ao Paulo, SP, 05314-970, Brazil}
\affiliation{Laborat\'orio Interinstitucional de e-Astronomia - LIneA, Rua Gal. Jos\'e Cristino 77, Rio de Janeiro, RJ - 20921-400, Brazil}
\author{M.~A.~G.~Maia}
\affiliation{Laborat\'orio Interinstitucional de e-Astronomia - LIneA, Rua Gal. Jos\'e Cristino 77, Rio de Janeiro, RJ - 20921-400, Brazil}
\affiliation{Observat\'orio Nacional, Rua Gal. Jos\'e Cristino 77, Rio de Janeiro, RJ - 20921-400, Brazil}
\author{J.~L.~Marshall}
\affiliation{George P. and Cynthia Woods Mitchell Institute for Fundamental Physics and Astronomy, and Department of Physics and Astronomy, Texas A\&M University, College Station, TX 77843,  USA}
\author{P.~Melchior}
\affiliation{Department of Astrophysical Sciences, Princeton University, Peyton Hall, Princeton, NJ 08544, USA}
\author{F.~Menanteau}
\affiliation{Department of Astronomy, University of Illinois at Urbana-Champaign, 1002 W. Green Street, Urbana, IL 61801, USA}
\affiliation{National Center for Supercomputing Applications, 1205 West Clark St., Urbana, IL 61801, USA}
\author{R.~Miquel}
\affiliation{Instituci\'o Catalana de Recerca i Estudis Avan\c{c}ats, E-08010 Barcelona, Spain}
\affiliation{Institut de F\'{\i}sica d'Altes Energies (IFAE), The Barcelona Institute of Science and Technology, Campus UAB, 08193 Bellaterra (Barcelona) Spain}
\author{F.~Paz-Chinch\'{o}n}
\affiliation{Department of Astronomy, University of Illinois at Urbana-Champaign, 1002 W. Green Street, Urbana, IL 61801, USA}
\affiliation{National Center for Supercomputing Applications, 1205 West Clark St., Urbana, IL 61801, USA}
\author{A.~A.~Plazas}
\affiliation{Department of Astrophysical Sciences, Princeton University, Peyton Hall, Princeton, NJ 08544, USA}
\author{E.~Sanchez}
\affiliation{Centro de Investigaciones Energ\'eticas, Medioambientales y Tecnol\'ogicas (CIEMAT), Madrid, Spain}
\author{V.~Scarpine}
\affiliation{Fermi National Accelerator Laboratory, P. O. Box 500, Batavia, IL 60510, USA}
\author{S.~Serrano}
\affiliation{Institut d'Estudis Espacials de Catalunya (IEEC), 08034 Barcelona, Spain}
\affiliation{Institute of Space Sciences (ICE, CSIC),  Campus UAB, Carrer de Can Magrans, s/n,  08193 Barcelona, Spain}
\author{I.~Sevilla-Noarbe}
\affiliation{Centro de Investigaciones Energ\'eticas, Medioambientales y Tecnol\'ogicas (CIEMAT), Madrid, Spain}
\author{M.~Smith}
\affiliation{School of Physics and Astronomy, University of Southampton,  Southampton, SO17 1BJ, UK}
\author{E.~Suchyta}
\affiliation{Computer Science and Mathematics Division, Oak Ridge National Laboratory, Oak Ridge, TN 37831}
\author{M.~E.~C.~Swanson}
\affiliation{National Center for Supercomputing Applications, 1205 West Clark St., Urbana, IL 61801, USA}
\author{G.~Tarle}
\affiliation{Department of Physics, University of Michigan, Ann Arbor, MI 48109, USA}
\author{V.~Vikram}
\affiliation{Argonne National Laboratory, 9700 South Cass Avenue, Lemont, IL 60439, USA}
\author{J.~Weller}
\affiliation{Excellence Cluster Origins, Boltzmannstr.\ 2, 85748 Garching, Germany}
\affiliation{Max Planck Institute for Extraterrestrial Physics, Giessenbachstrasse, 85748 Garching, Germany}
\affiliation{Universit\"ats-Sternwarte, Fakult\"at f\"ur Physik, Ludwig-Maximilians Universit\"at M\"unchen, Scheinerstr. 1, 81679 M\"unchen, Germany}

\collaboration{The DES Collaboration}
\noaffiliation

\date{\today}

\begin{abstract}
Weak lensing by large-scale structure is a powerful probe of cosmology and of the dark universe.
This cosmic shear technique relies on the accurate measurement of the shapes and redshifts of background galaxies and requires precise control of systematic errors.
The Monte Carlo Control Loops (MCCL) is a forward modelling method designed to tackle this problem.
It relies on the Ultra Fast Image Generator (UFig) to produce simulated images tuned to match the target data statistically, followed by calibrations and tolerance loops.
We present the first end-to-end application of this method, on the Dark Energy Survey (DES) Year 1 wide field imaging data.
We simultaneously measure the shear power spectrum $C_{\ell}$ and the redshift distribution $n(z)$ of the background galaxy sample.
The method includes maps of the systematic sources, Point Spread Function (PSF), an Approximate Bayesian Computation (ABC) inference of the simulation model parameters, a shear calibration scheme, and the fast estimation of the covariance matrix.
We find a close statistical agreement between the simulations and the DES Y1 data using an array of diagnostics.
In a non-tomographic setting, we derive a set of $C_\ell$ and $n(z)$ curves that encode the cosmic shear measurement, as well as the systematic uncertainty.
Following a blinding scheme, we measure the combination of $\Omega_m$, $\sigma_8$, and intrinsic alignment amplitude $A_{\rm{IA}}$, defined as $S_8D_{\rm{IA}} = \sigma_8(\Omega_m/0.3)^{0.5}D_{\rm{IA}}$, where $D_{\rm{IA}}=1-0.11(A_{\rm{IA}}-1)$.
We find $S_8D_{\rm{IA}}=0.895^{+0.054}_{-0.039}$,
where systematics are at the level of roughly 60\% of the statistical errors.
We discuss these results in the context of earlier cosmic shear analyses of the DES Y1 data.
Our findings indicate that this method and its fast runtime offer good prospects for cosmic shear measurements with future wide-field surveys.
\end{abstract}
\preprint{FERMILAB-PUB-19-247-AE}
\preprint{DES-2018-0362}

\maketitle

\section{Introduction}

Recent observations combining different cosmological probes have led to the establishment of the $\Lambda$CDM concordance model for cosmology.
One of these probes is cosmic shear, the measurement of spatial correlations in the apparent shape of background galaxies due to the weak gravitational lensing effect.
Since the first statistical detections of the effect \cite{Bacon2000detection,Kaiser2000cosmicshear,VanWaerbeke2000detection,Wittman2000detection}, there have been a large number of measurements with larger sample sizes and improved accuracies.
Recently, several wide-field surveys have reported cosmic shear measurements with unprecedented accuracies, such as the Kilo Degree Survey (KiDS) \cite{Hildebrandt2018kidsviking}, the Subaru HSC survey \citep{Hikage2018hsc} and the Dark Energy Survey (DES) \cite{des2017cosmlogical}.

A key requirement in cosmic shear measurements is the control of systematics both for the measurement of the shear correlation function and for the redshift distribution $n(z)$ of the galaxy sample used for the shape measurements.
To tackle this problem, a number of shape measurement methods have been proposed \cite{Zuntz2013im3shape,Conti2017calibration,Sheldon2017metacalibration,Miller2013lensfit} and have been reaching an increasing precision.
In parallel, various photometric redshift methods have been developed to derive galaxy redshifts from multi-band imaging data \cite{Hoyle2018redshift,Bonnett2016redshift,Bilicki2017photometric,Cavuoti2017cooperative,Sanchez2014photometric}.

Recently, the Monte-Carlo Control Loop \cite{Refregier2013mccl} (MCCL) method was proposed to tackle the shear and $n(z)$ measurement jointly.
It is based on a forward modelling approach using the Ultra Fast Image Generator (UFig) \cite{Berge2013ufig}.
In this method, image simulations are first tuned to agree statistically with the target data set and then used to calibrate the cosmic shear measurement and to quantify its systematic uncertainty.
The method was first tested at the 1-point \cite{Bruderer2015Calibrated} and 2-point level \cite{Bruderer2017shear} using simulations as mock observed data, which were also used to study the propagation of systematic effects onto the final cosmic shear measurement.
The MCCL method was also used to determine the redshift distribution of cosmological samples of galaxies \cite{Herbel2017redshift}.
Recently, the galaxy population model resulting from the MCCL method derived from broad-band imaging data was successfully compared to the Sloan Digital Sky Survey (SDSS) spectroscopic sample \cite{Fagioli2018sdss} and the narrow band imaging Physics of the Accelerating Universe Survey (PAUS) \cite{Tortorelli2018pau}.

In this paper, we present the first end-to-end cosmological analysis using the MCCL method, applied to the DES Year 1 (Y1) survey.
It constitutes a non-tomographic re-analysis of this data set with an independent approach.
We start from co-added images and perform object detection, Point Spread Function (PSF) modelling, shear calibration, $n(z)$ measurement, covariance matrix calculation, power spectra measurement and cosmological likelihood analysis.
At the heart of this approach is the simultaneous measurement of the shear angular power spectrum $C_\ell$ and the redshift distribution $n(z)$ of the galaxy sample.
After describing the method and its specific implementation for DES Y1, we present our results and cosmological constraints, and compare our results to the earlier DES Y1 analysis.
We follow a blinding scheme throughout our work.
Finally, we discuss the application of our method to future data sets, such as DES future releases.

This paper is organised as follows.
In Section~\ref{sec:MCCL_method}, we review the main features of the MCCL approach.
In Section~\ref{sec:loop1}, we describe how the simulations are matched to the data.
Our measurements of the weak lensing power spectrum and the redshift distribution of the lensed galaxies is presented in Section~\ref{sec:joint_cal}.
We present our cosmology constraints in Section~\ref{sec:cosmo_constraints} and conclusions in Section~\ref{sec:conclusions}.
The appendix describes our blinding scheme, the PSF modelling, the implementation of Approximate Bayesian Computation (ABC), and the internal tests on simulations.

\section{Monte-Carlo Control Loops methodology}
\label{sec:MCCL_method}

In \cite{Refregier2013mccl}, a framework called Monte Carlo Control Loops was presented as a method for making robust cosmological measurements.
The key principle of this approach is to heavily rely on realistic simulations and to analyse simulations in exactly the same way as is done for the observations.
In doing this, we are able to rigorously test all aspects of the measurement process in the regime used in the analysis.
The MCCL method divides the measurement process into three key steps that we identify as control loops.
In the first step, control loop 1, the simulations are tested against the data using a set of diagnostics to ensure that the simulations have a high fidelity to the data in the spanned space.
The forward model includes the intrinsic galaxy population and the Milky Way (stars and dust), as well as measurement features, such as the Point Spread Function (PSF) and noise properties of the images.
The result of this first step is a set of model configurations that agree with the data.
In the second step of the MCCL process, control loop 2, these simulations are used to calibrate the galaxy shear and redshift measurement sections of the pipeline.
In the third step, control loops 3.1 and 3.2, the robustness of these measurements is tested by taking excursions away from the fiducial simulation configurations that were used to calibrate the measurements.
As well as allowing us to perform a tolerance analysis, this exploration of measurement sensitivities also allows us to account for uncertainty stemming from systematic errors in a probabilistic way.
The shear power spectrum and redshift distribution measurements are then used for cosmological inference that accounts for both statistical and systematic errors.

The implementation of MCCL in this work follows these steps in order to obtain the final cosmology constraint:
\begin{enumerate}[label=(\roman*), leftmargin=*]
    \setlength\itemsep{0.01em}
    \item we build parametric models for simulating co-added DES images including systematic maps, Milky Way and galaxy populations,
    \item we find a posterior on the model parameters using ABC,
    \item we run an ensemble of simulations of the full DES area using the points from the ABC posterior,
    \item we calculate shear calibration parameters and redshift distribution for each simulation,
    \item we apply the calculated shear calibration parameters to the galaxy catalogues obtained from the DES images to create a family of pairs of $C_{\ell}$ and $n(z)$ corresponding to the ABC posterior,
    \item we calculate cosmology constraints for each pair of $C_{\ell}$ and $n(z)$,
    \item we combine the ensemble of cosmological constraints to create the final constraint that marginalises over shear calibration and redshift distribution uncertainties.
\end{enumerate}
This pipeline was accompanied by an array of tests, such as recovery of input $C_{\ell}$ from simulations, the impact of model extensions, and discrepancies in systematic maps.
These tests are described in the sections below.
Step (i) above corresponds to control loop 1, step (iv) to control loop 2, step (ii) to control loop 3.1, while testing the model extensions in Section~\ref{sec:loop32} to loop 3.2.
We follow a blinding scheme and define a set of conditions to be met before unblinding in Section~\ref{sec:blinding}.

\section{Fiducial simulation parameters (Loop 1) }
\label{sec:loop1}

As stated earlier, the MCCL approach implemented in this work relies on modelling of all important features that have an impact on the key measurements of shear and redshifts of galaxies.
These include the intrinsic properties of the galaxy population over cosmic time, a model of the Milky Way, and observational features linked to the data taking.
In this section, we present a brief description of these components along with our measurement and results that lead to our fiducial simulation parameters for the later work.

\subsection{Galaxy population model}
\label{subsec:galaxy-population-model}

A detailed description of the features of the intrinsic galaxy population model used in this study is given in \cite{Herbel2017redshift}.
In order to render images of galaxies, we need to assign fluxes, light profiles, and positions to each galaxy.
We do this by first modelling the galaxy luminosity distributions of different galaxy populations, red and blue, using Schechter functions $\phi$, which can evolve with redshift.
By drawing from these functions, we are able to generate a sample of galaxies with redshifts and absolute magnitudes.
Next, we draw a rest-frame spectral energy distribution (SED) for each galaxy.
We model the SEDs as a linear combination of five template spectra, which are based on the Bruzual-Charlot stellar evolution synthesis models \cite{Bruzual2003} and which are also used by \textsc{kcorrect} \cite{Blanton2006}.
The corresponding coefficients are sampled from a Dirichlet distribution, which is motivated empirically by data from the SDSS, as described in \cite{Herbel2017redshift}.
At this step, we again make a distinction between the red and the blue galaxy populations by using two distinct Dirichlet distributions.
After each galaxy has been assigned a spectrum, we are able to compute apparent fluxes in arbitrary filter bands, which are used to render the objects on our simulated images.
We also include reddening by Galactic dust using the extinction map derived by \cite{Schlegel1998}.
The positions of galaxies on the sky are drawn uniformly, without clustering.

After assigning fluxes to our galaxies, we randomly draw a light profile for each object.
We use S\'ersic profiles \cite{Sersic1963} parameterized by the S\'ersic index $n$ and the half-light radius $r_{50}$ to model the light distributions of our galaxies.
To assign half-light radii, we use the model given by \cite{Herbel2017redshift}, i.e., we sample physical half-light radii for our galaxies from a log-normal distribution with a fixed standard deviation and a mean that depends on the absolute magnitudes of the galaxies.
We then transform the physical size to an angular size on the sky using the angular diameter distance, calculated using the fiducial cosmological model.

We assign the same S\'ersic index $n_\text{blue}$ to all galaxies sampled from the blue population and the same S\'ersic index $n_\text{red}$ to all galaxies sampled from the red population, whereby $n_\text{blue} \neq n_\text{red}$.
This is motivated by results from the literature where it was found that blue galaxies are on average well described by a S\'ersic index $n = 1$ and red galaxies are well matched using $n = 4$ \cite{Tarsitano2018morphology,Leauthaud2012evolution}.
The value of $n_\text{blue}$ is found using the ABC scheme, while $n_\text{red}=4$ is fixed (see Section~\ref{sec:abc-fits}).

Finally, each simulated galaxy is assigned an intrinsic ellipticity described by two components $e = (e_1,e_2)$.
We do it by drawing an ellipticity magnitude $|e|$ from the $p(|e|)$ distribution and rotate it by a random angle.
We use a $p(|e|)$ based on the Beta distribution.
Our model uses two parameters: $e_{\rm{ratio}}$ and $e_{\rm{sum}}$, which map to Beta distribution parameters $\alpha, \beta$ in the following way:
$\alpha=e_{\rm{sum}}e_{\rm{ratio}}$ and  $\beta=e_{\rm{sum}}(1-e_{\rm{ratio}})$.
Variation in $e_{\rm{ratio}}$ corresponds to shifting the mode of the distributon between 0 and 1.
Value of $e_{\rm{sum}}$ close to zero results in an distribution that is close to uniform, while large $e_{\rm{sum}}$ in a narrow spread around the mode.
The prior is on these parameters is specified in Appendix~\ref{sec:variable-parameters-priors}.
The posterior is found using the ABC scheme, see Figure \ref{fig:abc_posterior_samples}.

We assume a cosmological model to calculate the angular diameter distances in the calculation of magnitudes and sizes of galaxies.
We use the same cosmological parameters as in \citep{Herbel2017redshift}.
As the ABC posterior is tuned to data and constrained by magnitudes, sizes, and colours of the galaxies detected in the images, as well as the spectroscopic redshift sample from VVDS.
Using a slightly different cosmology parameters would modify angular diameter and luminosity distances, and these changes would be, to first order, compensated by modifying other model parameters, such as the normalisation or redshift evolution of the luminosity and size functions.
As these parameters are degenerate and the posterior is anchored on the imaging and spectroscopic data, we do not expect the calculated $n(z)$ and shear calibration to change significantly.
Therefore, we do not expect this assumption to influence the cosmological constraints measured in this work.
It may prove useful to investigate this dependence in more detail for future lensing surveys.

\subsection{Milky Way model}

To generate a catalogue of stars for rendering the simulated image, we combine the stars in the \textit{Gaia} Data Release 2 (DR2)  \citep{Gaia2018dr2}, with the Besan\c{c}on model \footnote{http://model.obs-besancon.fr} of the Milky Way \citep{Robin2003} (see Section \ref{sec:systematics_maps}).
The \textit{Gaia} objects are placed on the image according to their actual position on the sky, such that we estimate the PSF in the simulations at the same positions as in the data.
To generate the faint end of the stellar population, we use the Besan\c{c}on model, which is based on stellar population synthesis.
We evaluate the model for all HEALPix \citep{Gorski1999healpix} pixels of a map with \texttt{nside}=8 that overlap with the DES Y1 area.
We create Besan\c{c}on catalogues that cover an area of $5\,\text{deg}^2$ and subsample these catalogues according to the area covered by the simulated images.
This way the variation in density is included in the simulations.

To combine the stars from \textit{Gaia} with the ones generated by the Besan\c{c}on model, we map the apparent CFHT-MegaCam magnitudes of the Besan\c{c}on stars to the \textit{Gaia} $G$-band using the relation provided in \citep{Jordi2010gaia} (second equation in section 5.2 and table 7).
For each \textit{Gaia} object, we then find the closest match from the Besan\c{c}on stars in terms of the $G$-band apparent magnitude.
The matched Besan\c{c}on stars are subsequently placed at the positions of the corresponding \textit{Gaia} objects.
While not all objects in the \textit{Gaia} catalogue are true stars, at this stage we do not attempt to improve the purity of the sample.
This is, however, addressed at the PSF modelling step (See Appendix~\ref{appendix:psf}).

\subsection{Model of the measurement process}
\label{sec:systematics_maps}

We analyse co-added DES images, as well as simulate co-adds with UFig.
We do not analyse single exposure images in our method, although we use information about them to create systematic maps of the PSF and noise in the co-adds.
Each exposure taken by the Dark Energy Camera \citep{Flaugher2015decam} comprises of 62 images, each taken by a single chip.
Additional 12 chips are used for guiding and focus.
However, the Y1 images were constructed mostly from 67 CCDs due to various instrumental issues \citep{DrlicaWagner2018y1gold}.
In the DES pipeline, each co-add image is created by adding single chip images from multiple exposures.
Before the co-addition, the chip images are resampled to the co-add coordinate system using an astrometric solution \cite{Morganson2018pipeline,Bernstein2017astrometric}.
Therefore, the co-add image properties, such as noise levels or the PSF, can sharply change across the image in places corresponding to boarders of single chip images.
To include this effect in simulations, we create a set of Boolean maps for each co-add, which contains information about each exposure's contribution to each pixel in the co-add.
We create the Boolean exposure maps for all $grizY$ bands and use them to create noise level and PSF maps.

\begin{figure}[t!]
\includegraphics[width=1.0\linewidth]{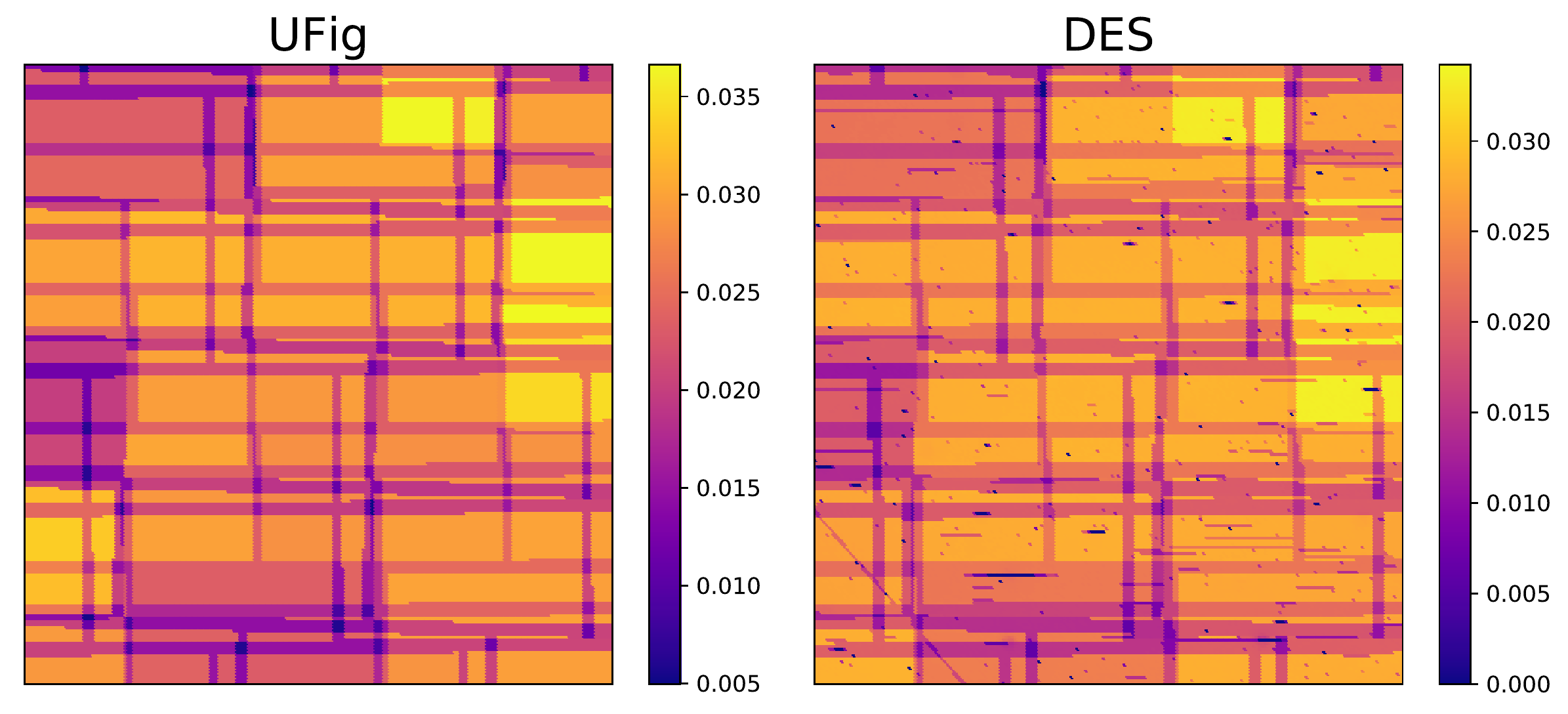}
\caption{Example noise map for tile DES0622-6039.
The colour shows the inverse variance of the pixel noise.
The tiling pattern was constructed using the astrometric information of every chip that was used for constructing the COADD.}
\label{fig:example_noise_map}
\end{figure}

Noise level maps contain information about the noise standard deviation for each pixel in the co-add.
They are created using the noise level estimate in the headers of single chip images, based on the \verb SKYSIGMA ~field.
This field contains the standard deviation of the sky background noise.
A weighted estimate is created for the co-add pixels with multiple single image contributions using the same weighting scheme as used in the co-add production process.
An example noise map is shown in Figure~\ref{fig:example_noise_map}.
To fine-tune the noise level, the map is then multiplied by a scaling parameter $s_\mathrm{bkg}$, which is found using ABC (see Appendix~\ref{appendix:abc-analysis}).
For each simulated image, we draw the Gaussian noise realisation from the noise map.
To emulate the effects of the co-addition process on the images, we convolve the drawn noise with a specially designed kernel, which is created so that the auto-correlation of the convolved noise image is similar to that expected from the Lanczos resampling with n=3, as employed by the DES pipeline \citep{Morganson2018pipeline} (see Appendix~\ref{appendix:noise_correlation} for details).

Objects in the images are detected by \textsc{SExtractor}~\citep{Bertin1996SExtractor}.
We analyse DES and UFig images using the same \textsc{SExtractor} settings (see Appendix \ref{appendix:galaxy_sample_selection}).
While running \textsc{SExtractor} on DES data, we used the noise maps accompanying the co-add images and produced by the DES pipeline.
For simulations, the noise maps were taken from the inverse variance maps described above.
We verified that there is no significant difference on measured moments when using one or the other map, as \textsc{SExtractor} rescales the noise maps internally after performing noise level estimation.

In the DES pipeline, the background is subtracted from each single exposure before co-addition.
We simulate the co-adds directly according to noise and PSF maps, and do not include a background light model.
We do, however, subtract the global mean of the image.
To address this slight discrepancy, we include both global and local background subtraction in our \textsc{SExtractor} runs on the DES images and simulations.
We verified that the \textsc{SExtractor} output is robust to the level of background on the images for our data.

\begin{figure}[t!]
\includegraphics[width=1.0\linewidth]{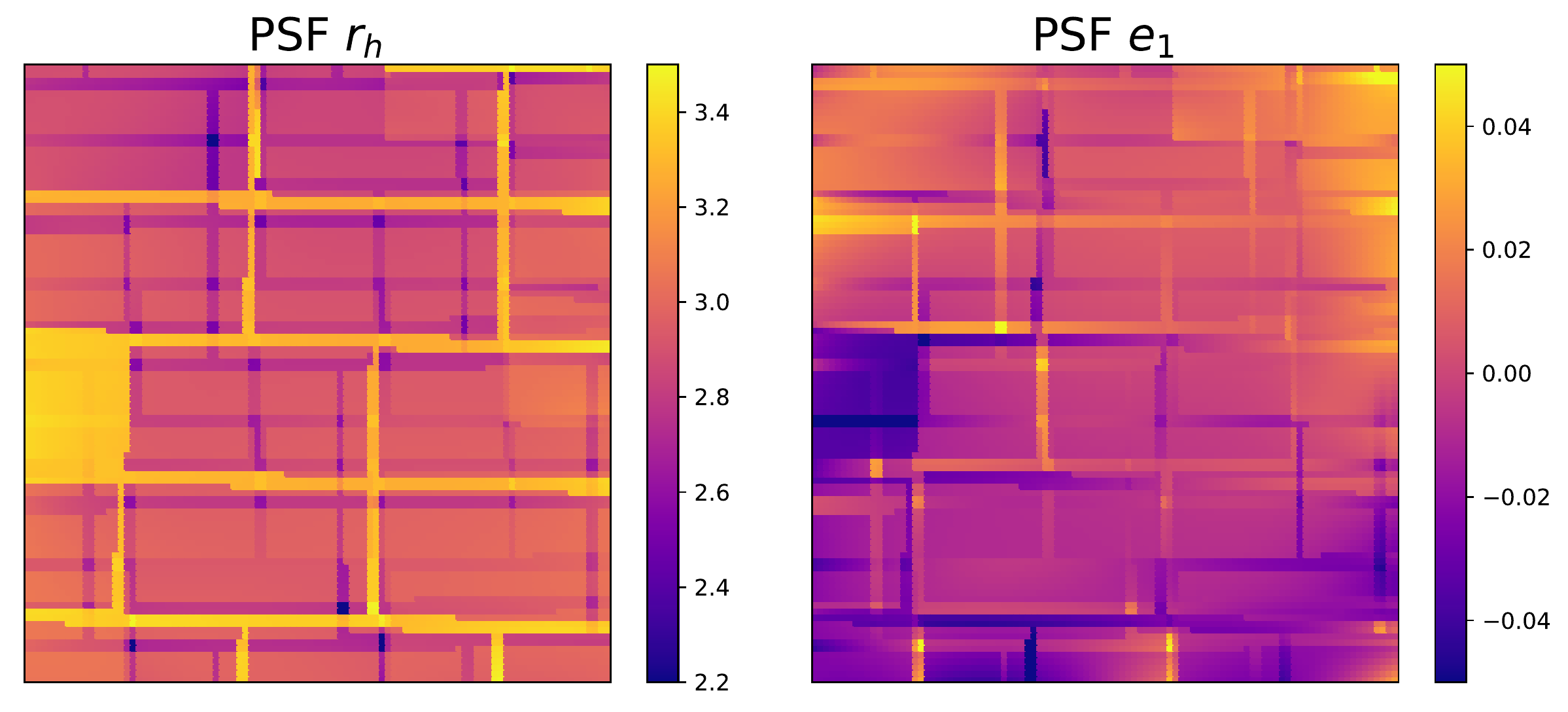}
\caption{Example PSF maps for tile DES0622-6039.
Maps for PSF FWHM $r_p$ and ellipticity $e_1$ are shown in the left and right panels, respectively.
Maps for flexions, kurtosis and flux ratio from two Moffat profiles were also created in the same fashion.}
\label{fig:example_PSF_map}
\end{figure}

\begin{figure*}[t!]
\includegraphics[width=1\linewidth]{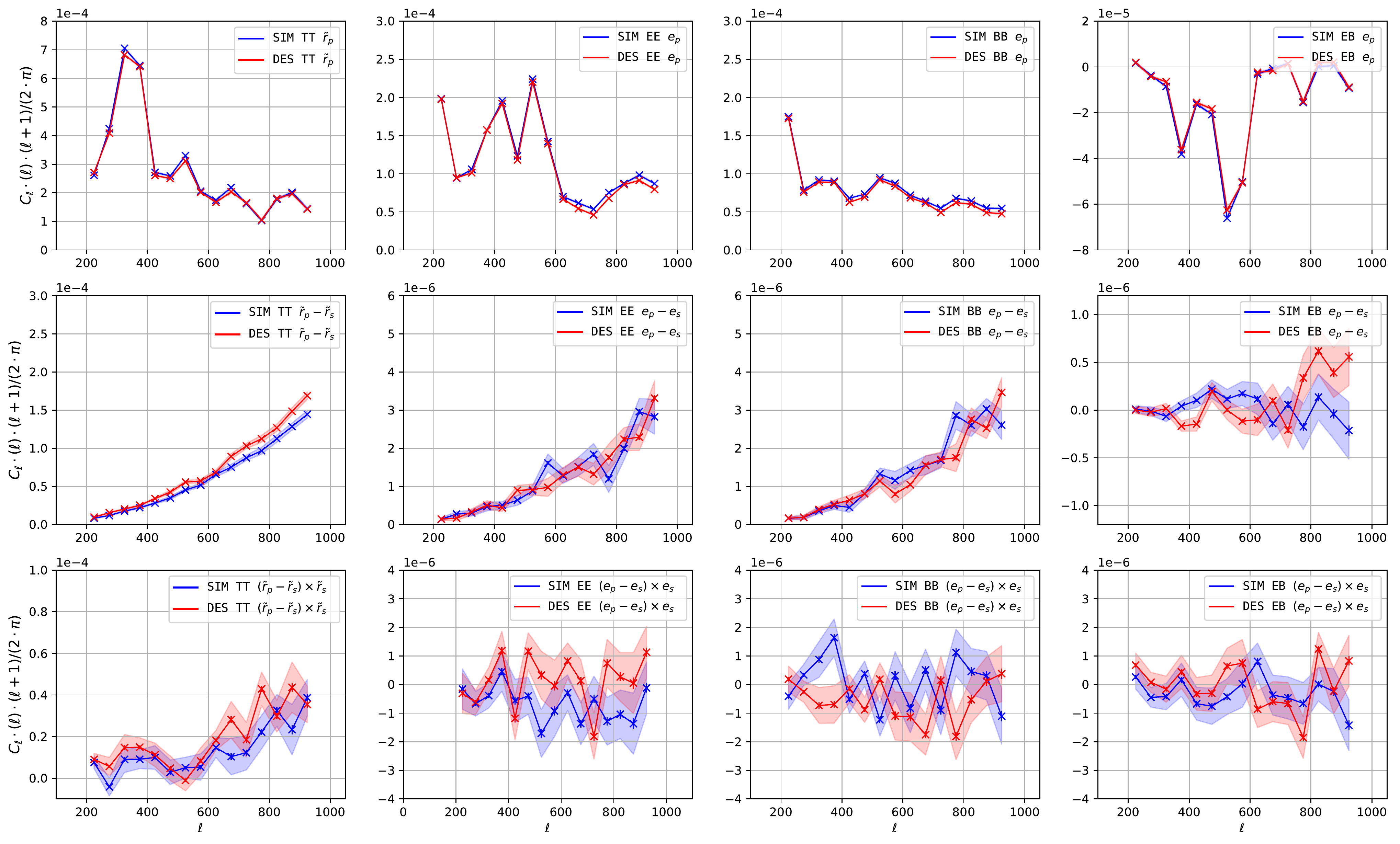}
\caption{Agreement between the PSF 2-pt functions in the DES data (red lines) and UFig simulations (blue lines).
The upper left panel shows the TT power spectrum for PSF FWHM $\tilde r_p$, calculated in fractional deviation form $\tilde r=(r-\bar r)/\bar r$, where $\bar r$ is the mean size.
The remaining upper panels show EE, BB, and BE spectra for shape parameters $e_p$ of the PSF model calculated at the positions of galaxies.
The spectra of the residuals $\tilde r_p - \tilde r_s$ and $e_s-e_p$ between the measured sizes and shapes of validation stars $\tilde r_s, e_s$ and PSF model $\tilde r_p, e_p$ at positions of stars is shown in the middle panels.
Cross-spectrum between the residuals $\tilde r_p - \tilde r_s$, $e_s-e_p$, and PSF esimates $\tilde r_s, e_s$, are shown in the bottom panels.
Middle and lower panels show the harmonic space equivalent of $D_1$ and $D_2$ statistics proposed in \citep{Rowe2010psf} as diagnostic tools for PSF model selection.
The error-bands correspond to 1$\sigma$ standard deviation calculated from multiple realisations of the fiducial simulation with different random seeds.
}
\label{fig:psf_2pt_agreement}
\end{figure*}

The PSF model is based on three key elements: parametric PSF models, fast parameter measurement with deep learning, and interpolation on co-adds.
This pipeline is independent of that used in \citep{Zuntz2017catalogues}, and allows for fast modelling inside the control loops.
We measure the PSF parameters only from the objects identified in the \textit{Gaia} catalogues, with magnitudes in the DES $r$-band $17<m<22$.
The PSF model is based on a double Moffat \citep{Moffat1969psf} profile, with $\beta_1=2$ and $\beta_2=5$.
It has 9 parameters: size, ellipticity (2x), flexions (4x), kurtosis, and the ratio of fluxes between the two Moffat profiles.
We obtain the parameters of that model using a deep learning method described in \citep{Herbel2018Fast}, with few modifications (see Appendix \ref{appendix:psf}).
PSF maps are created using the Boolean exposure maps described above.
We interpolate these parameters across the co-add plane using a basis that combines Chebyshev polynomials and the information from the Boolean exposure maps (see Appendix \ref{appendix:psf} for more details).
This way, the discontinuities in the PSF variations across the co-add can be included.
We use a robust fitting algorithm with a $\sigma$-clipping procedure, which aims to remove unusual stars, including false positives in the \textit{Gaia} catalogue.
For each tile, a randomly chosen set of 15\% of the stars are excluded from being used as an input to PSF model fitting.
These stars constitute a validation sample, which is used to calculate residuals between the interpolated PSF and measured star parameters.
Figure \ref{fig:example_PSF_map} shows an example PSF map for the PSF size and ellipticity parameters.
These models are used for making the forward simulations, as well as for the shear measurement.

We simulated the full DES area using our forward model and used the exact same set of DES and UFig tiles to perform our analysis.
Figure \ref{fig:psf_2pt_agreement} shows the agreement between simulations and DES data in terms of PSF power spectra.
The power spectra were calculated using \textsc{PolSpice} and described in Appendix~\ref{appendix:power_spectrum_calculation}.
The parameters of the PSF, calculated at the positions of galaxies, are: PSF FWHM $r_p$ and ellipticity $e_p$ (top panels).
Middle panels show the power spectra of the residual between the PSF estimates, $r_p, e_p$, and the measurement from validation stars $r_s, e_s$.
Bottom panels show the cross power spectrum between the PSF model and the residual $e_s-e_p$, $r_s-r_p$, at the positions of validation stars.
The residual auto and residual $\times$ model power spectra were noise-corrected.
The bands correspond to 1$\sigma$ standard deviation and are calculated from multiple realisations of the fiducial survey with different random seeds.
The agreement is generally very good for the PSF parameters and the residual power spectra.
Small discrepancy is observed in the PSF size residual auto power spectrum.
Model $\times$ residual spectra are very low and also in agreement.
The PSF power spectra for different simulations varied slightly due to different star sample selection and their measured parameters.
This was caused by random selection of validation stars, pixel noise, which affected the star parameter measurement, and blending of the PSF stars with other objects.
The differences between these power spectra were, however, very small, and we do not show them here.

\subsection{ABC fits to DES data}
\label{sec:abc-fits}

\begin{figure*}[t!]
\includegraphics[width=1.0\textwidth]{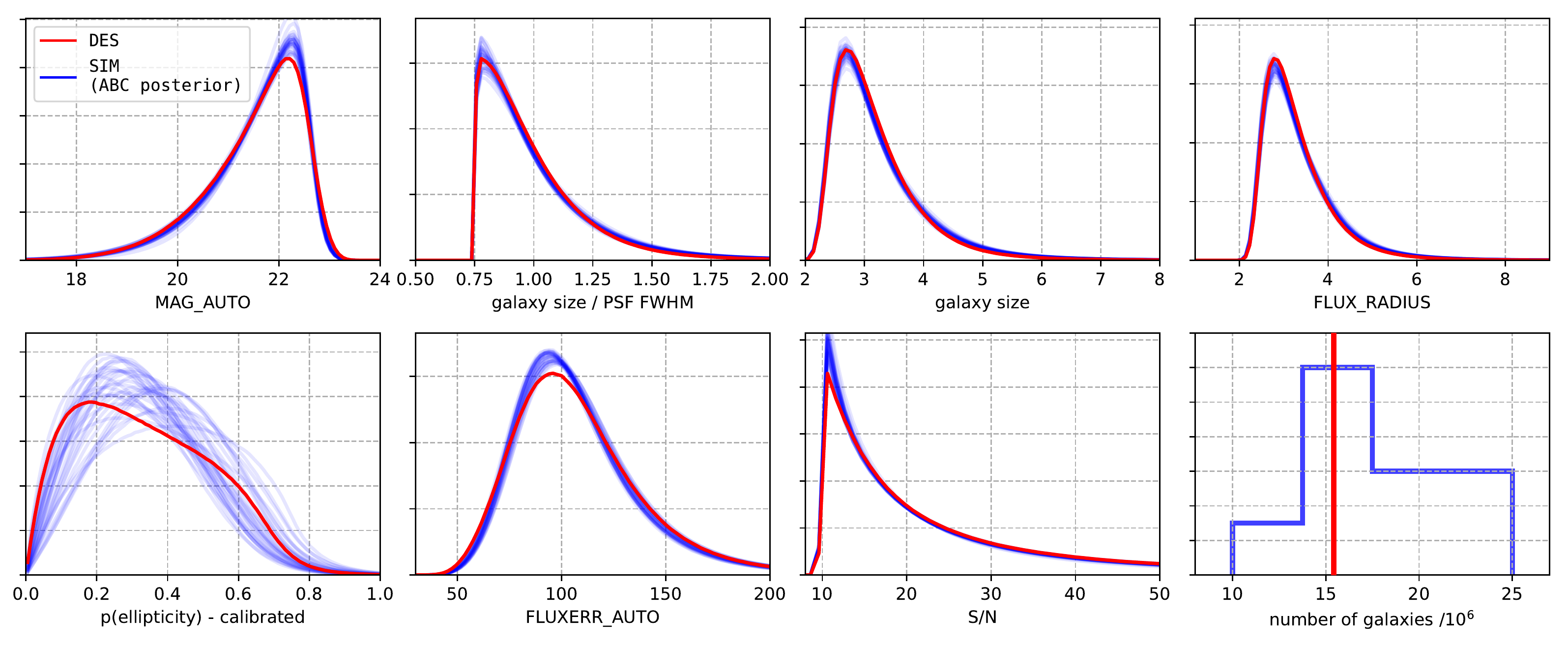}
\caption{Agreement between ABC posterior and DES data.
The red line shows the histograms from DES data and the blue lines from the \nsurveys\ UFig simulations from ABC posterior.
These normalised histograms were created using the full DES catalogue, as well as catalogues created from simulating the full area.
The bottom right panel shows the number of galaxies in the DES data (red line) and \nsurveys\ UFig simulations (blue dots).
The y-axis corresponds to the indices of \nsurveys\ simulations.
The galaxy size was calculated using Equation~\ref{eqn:galaxy_size}.
The Signal-to-noise ratio (S/N) has been calculated from \textsc{SExtractor} parameters as \texttt{FLUX\_AUTO}/\texttt{FLUXERR\_AUTO}.
}
\label{fig:histograms_des_abc}
\end{figure*}

We follow the method detailed in \cite{Herbel2017redshift} to generate a family of image simulations that are statistically consistent with the DES Y1 data.
The method we use to adjust our model to the survey data is called Approximate Bayesian Computation \cite{Sunnaker2013, Akeret2015}.
It allows for Bayesian inference in situations where the likelihood function is not tractable, which is the case for our simulations:
there is no clear empirical expression for the likelihood, neither on the image nor on the catalogue level.
However, since we are able to compare the simulations to survey data using distance metrics, the ABC framework allows us to approximate the corresponding true Bayesian posterior.

The parameter space we sample during the ABC analysis has 35 dimensions.
We vary six sets of parameters:
($i$) the parameters controlling the redshift evolution of the luminosity functions,
($ii$) the parameters of the Dirichlet distributions used to sample galaxy SEDs,
($iii$) the parameters of our model for the intrinsic size of galaxies,
($iv$) the value of the S\'ersic index $n_\text{blue}$ for galaxies sampled from the blue population,
($v$) the parameters controlling the distribution from which we sample intrinsic galaxy ellipticities,
($vi$) a parameter scaling the background level of our simulated images (see Table~\ref{tab:abc-parameters} for more details).
We choose these parameters because they have the largest impact on the posterior $n(z)$ curves and on the shear calibration.
The S\'ersic index for red galaxies $n_\text{red}$ is fixed because it impacts $n(z)$ and the shear calibration only very weakly.
Furthermore, since there are only few red galaxies compared to the blue population, we have little constraining power on this parameter.
In Appendix \ref{sec:variable-parameters-priors}, we give more information on our parameter space and specify the priors.

The distance metrics we use probe basic properties of the simulated images such as number counts, the distribution of measured galaxy magnitudes, sizes and ellipticities as well as galaxy colours.
Furthermore, we include spectroscopic data from the VIMOS VLT Deep Survey (VVDS \cite{LeFevre2005, Garilli2008, LeFevre2013}) to tighten the constraints on $n(z)$.
In total, we use a combination of five distance metrics to obtain a posterior; further information on this is given in Appendix \ref{sec:distance-metrics}.
To compute the distance values for one sample, we evaluate our model on 20 randomly chosen DES tiles, which corresponds to an area of $10.7\,\text{deg}^2$ (we use the same tiles for all samples). We then compute distance metrics tile-by-tile and average the resulting values to reduce the impact of cosmic variance. In total, we evaluate our model for $110\,000$ prior samples.

In Appendix \ref{appendix:abc-analysis}, we show the ABC posterior obtained from the analysis described above.
We do not show the parameters controlling the coefficient distributions used to assign spectra to galaxies, since we have little constraining power on these parameters, so that we effectively marginalize over them.
Furthermore, we compare histograms of various galaxy quantities measured from the DES data and from the posterior simulations in Figure \ref{fig:histograms_des_abc}.
We find that the DES histograms (red line) lie within the histograms measured from UFig simulations of ABC posterior (blue lines).
That is the case for the bulk of the distributions, some small discrepancy is visible in the tails.
Small discrepancy is visible in the \texttt{FLUXERR\_AUTO} parameter, but the overall shape of the curves match well.
The overall agreement ensures that the DES data lies within the simulation space, according to our metrics.
The uncertainty on the overall number of galaxies and the $p(e)$ is larger than for the other parameters.
It can be improved in future work by running the ABC algorithm for longer until full convergence.
However, it is not necessary to decrease this uncertainty further at this point, as the ABC posterior is always a conservative approximation to the true posterior.
Lack of convergence results in a larger systematic uncertainty on $n(z)$ and shear calibration and propagates to the cosmological parameters.
The uncertainty in the number of galaxies does not affect the covariance matrix for the DES power spectrum, as it is calculated using the DES shapes directly, as described in Section~\ref{sec:covmat}.

We choose the sample with the lowest combined distance measure as our fiducial parameter set.
This set is then used to create the fiducial simulation, on which many of the following basic tests are performed.

\section{Joint Shear Power Spectrum and Redshift Measurement (Loop 2 and 3)}
\label{sec:joint_cal}

In the MCCL method, we jointly measure the redshift distribution of source galaxies $n(z)$ and the shear power spectrum $C_{\ell}$.
To achieve this, we use exactly the same simulated galaxy catalogue from UFig simulations to calculate an $n(z)$ distribution and shear calibration parameters.
Moreover, by using a set of surveys from the ABC posterior that are compatible with the DES data, we effectively quantify the uncertainty on $n(z)$ and the shear calibration.

Our method is designed to measure the shear power spectrum and $n(z)$ only \citep{Bruderer2017shear,Herbel2017redshift}.
The shear estimates are not designed to be robust to all systematics; in fact, the measurement relies heavily on calibration for a dataset with specific properties.
The shear bias as a function of various quantities, like signal-to-noise ratio or galaxy size, may remain non-zero \citep{Bruderer2017shear}.
This way, the calibrated shear catalogue can be considered only to be an intermediate product.
The confidence about the accuracy of the results stems from the fact that the simulations are well matched to the observations and display similar biases, and that we correctly recover the shear power spectrum in simulations.
As long as similar biases are present between the DES and UFig data, and the power spectrum is recovered correctly in the simulations, the DES measurement should is expected to be equally accurate.

We simulate \nsurveys\ UFig surveys of the full Y1 area using the points from the ABC posterior, including the fiducial survey.
Each of these \nsurveys\ simulations is used to calculate the redshift distribution and a set of shear calibration parameters.
These parameters are then applied to the fiducial UFig catalogue and to the DES catalogue, and the power spectra are computed.
This way, we obtain \nsurveys\ $C_{\ell}$ and $n(z)$ pairs, for both DES and the fiducial UFig survey.
Variations in these parameters capture the uncertainty in shear and redshift inside the ABC posterior.

\subsection{Shear calibration and power spectrum measurement}
\label{sec:shear_calibration}

We follow the method presented in \citep{Bruderer2017shear,Bruderer2015Calibrated} with several modifications.
The method in \citep{Bruderer2017shear,Bruderer2015Calibrated} uses quantities measured by \textsc{SExtractor} \citep{Bertin1996SExtractor} and PSF parameters to create the shear estimator for each galaxy.
In this work, we use the PSF parameters outputted by the Convolutional Neural Network (CNN), described in \citep{Herbel2018Fast}.
Further modifications include the correction of the effect of the \textsc{SExtractor} weight function used to measure the quadrupole moments.
For the details of \textsc{SExtractor} run, see Appendix \ref{appendix:galaxy_sample_selection}.
Then, we create shear maps using the HEALPix pixelisation scheme \citep{Gorski1999healpix} and measure their power spectrum with \textsc{PolSpice} \citep{Szapudi:2001, Szapudi:2001ab, Chon:2004}.
This process is described in detail in Appendix~\ref{appendix:power_spectrum_calculation}.

\subsubsection{Shear measurement and calibration}

We use \textsc{SExtractor} weighted moments \verb+X2WIN_IMAGE+, \verb+Y2WIN_IMAGE+ and \verb+XYWIN_IMAGE+ to create the moment matrix $\mathbf{M}$.
Similarly, we use the PSF size and ellipticity to create the PSF moment matrix $\mathbf{P}$.
To measure the weighted moment, \textsc{SExtractor} uses a Gaussian weight function with width $\sigma_w = 0.5 \cdot \verb+FLUX_RADIUS+^2/ \log(2)$, where $\verb+FLUX_RADIUS+$ is the measured half-light radius  \citep{Bertin1996SExtractor}.
The estimated, deconvolved galaxy moment is then
\begin{equation}
    \mathbf{Q} = \left[ \mathbf{M}^{-1} - \alpha_2 \mathbf{W}^{-1} \right]^{-1} - \alpha_1 \mathbf{P},
    \label{eqn:shear_calibration}
\end{equation}
where $\mathbf{W}$ is diagonal with $W_{ii}=\sigma_w$.
Parameters $\alpha_1$, $\alpha_2$ and $\eta$ (defined below) control the shear calibration and are found using simulations.
This equation gives the correct, deconvolved moment for $\alpha_1=\alpha_2=1$ if the observed galaxy, the PSF, and the weight are all Gaussian.
The shear estimators are given by
\begin{align}
    \gamma_1 =& \eta \frac{Q_{11}-Q_{22}}{Q_{11}+Q_{22}}, \\
    \gamma_2 =& \eta \frac{2Q_{12}}{Q_{11}+Q_{22}}.
\end{align}
The calibration parameter $\alpha_2$ was set to $\alpha_2=1$.
We vary $\alpha_1$ on 200 grid points between $\alpha_1 \in [0.6, 0.8]$ and select the value that minimizes the PSF leakage, for PSF ellipticity binned in 50 bins of equal size between $e_p^{\rm{min}}=-0.1$ and $e_p^{\rm{max}}=0.1$.
For each value of $\alpha_1$ we choose $\eta$ that minimizes the shear multiplicative bias for these bins.
For the fiducial simulation, the calibration parameters were: $\eta=0.7616$ and $\alpha_1=0.7246$.
Across the 30 simulations from the ABC posterior, we found
$\eta =0.7742 \pm 0.0222$ and
$\alpha_1= 0.7280 \pm 0.0050$.

The source galaxy sample was created by applying a range of cuts on galaxy and PSF size ratio, signal-to-noise, \textsc{SExtractor} flags, maximum ellipticity, and others, as described in Appendix~\ref{appendix:galaxy_sample_selection}.
The catalogues are created using \band-band objects only, and contain
15,432,057 objects for DES data and
15,370,564 for fiducial UFig survey,
and vary slightly when different calibration parameters are applied.
This corresponds $\approx$3 galaxies/arcmin$^2$.

\subsubsection{Systematics model}
\label{sec:systematics_model}

\begin{figure*}[t!]
\includegraphics[width=1\textwidth]{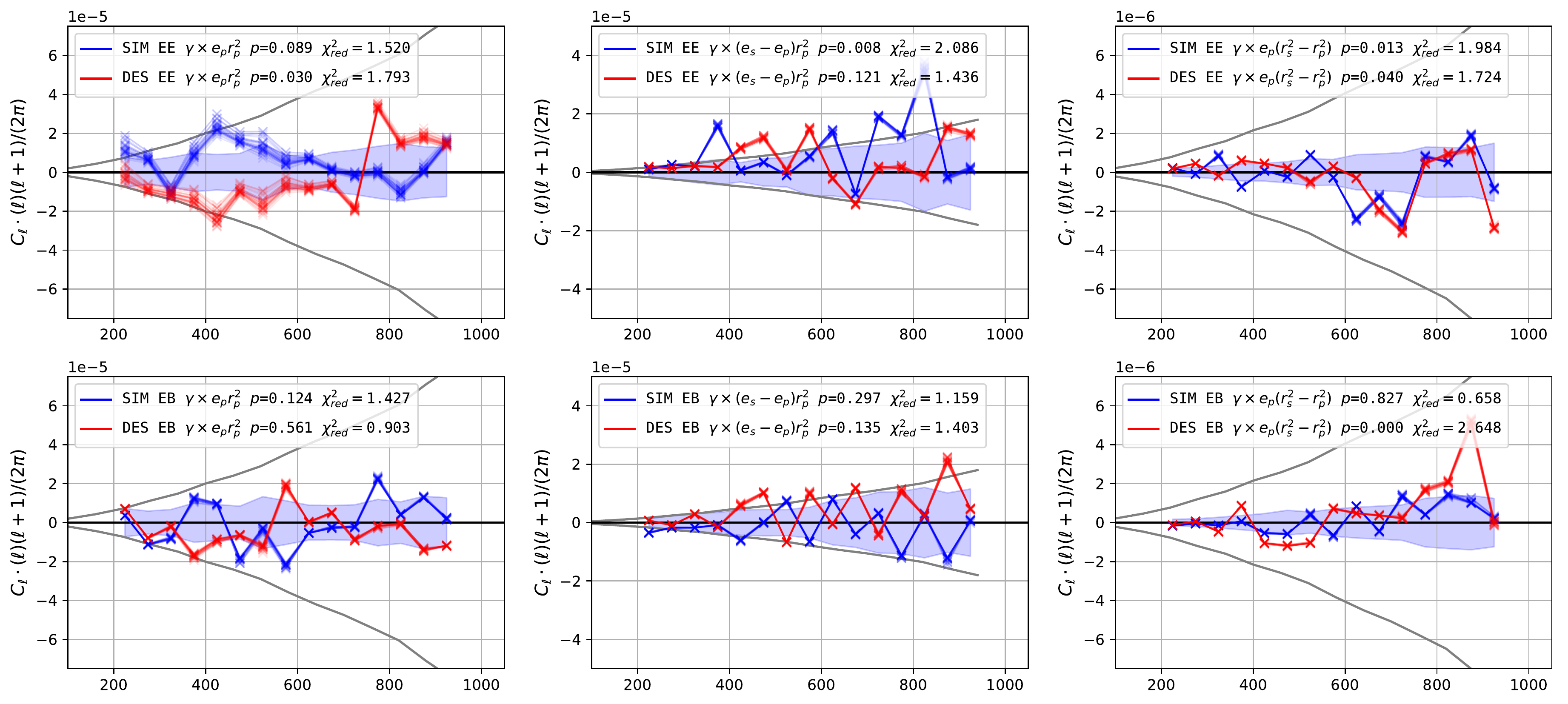}
\caption{Diagnostic 2-pt functions of the DES data.
The panels show the cross power spectrum between the shear E-mode and PSF model parameters: ellipticity $e_p$ and FWHM $r_p$ at the galaxy position, as well as the residual between the PSF model and star measurement $e_s-e_p$ and $r_s^2-r_p^2$.
Red and blue lines correspond to DES and fiducial UFig simulation, respectively, calibrated with \nsurveys\ calibration parameter sets from the ABC posterior.
The blue errors bands correspond to the statistical uncertainty $\sigma_{\mathrm{sys}}^2$ and is calculated from the standard deviation of $C_{\ell}$ calculated from \nsurveys\ simulations of the fiducial survey with different random seeds.
The $\chi^2$ neglects the covariance between the elements of the $C_{\ell}$ vector.
The number of degrees of freedom is $N_{\rm{dof}}=15$.
The grey lines correspond to the requirements described in Section~\ref{sec:systematics_model}.
In order to shift the cosmology contours by 0.5$\sigma$, all $C_{\ell}$ elements would have to consistently exceed the requirement.
The requirements are conservative, as they were based on statistical uncertainty only, and not including systematic errors from shear calibration and $n(z)$.
}
\label{fig:shear_cross_PSF}
\end{figure*}

We assess the quality of the shear measurement by examining the 2-pt statistics: the B-modes and shear-PSF cross power spectra.
We additionally investigate shear 1-pt statistics by looking at the mean shear as a function of PSF and galaxy parameters.
We find small PSF leakage and significant mean shear for the $\gamma_1$ component.
We do not expect to significantly affect the 2-pt measurement, as discussed in Section~\ref{sec:remaining_issues} and Appendix~\ref{appendix:shear_1pt}.

We model the systematic contributions to the measured ellipticity $\gamma_{\rm{obs}}$ from the PSF in the following way:
\begin{equation}
\label{eqn:systematic_model}
\gamma_{\rm{obs}}^{i} = \gamma_{\rm{true}} (1+ \beta_m^{i} \delta r_{p} )  + \alpha^{i} e_{p}^{i} r_{p}^2 + \beta^{i}_e \delta e_{p}^{i} r_{p}^2 + \beta^{i}_r e_{p}^{i} \delta r_{p}^2
\end{equation}
where $r_p$ is the PSF size, $e_{p}^{i}$ is the PSF shape and $\delta r_{p}$, $\delta e_{p}^{i}$ are the errors in the PSF model for size and shape, respectively.
Coefficient $\alpha^i$ quantifies the effect of the error in PSF deconvolution.
Coefficients $\beta_e$, $\beta_r$ capture the effect of errors in the PSF model.
Coefficient $\beta_m$ is responsible for multiplicative bias arising from the error in PSF size model.
This model is loosely based on the linearised error propagation model in \citep{PaulinHenriksson2008psf} and extends the model used in \citep{Hildebrandt2016kids} by adding an additional scaling by PSF size.
We can estimate $\alpha$ directly from the data by measuring the slope of $\gamma_{\rm{obs}}$ as a function of $e_{p}^{i} r_{p}^2$.
Coefficients $\beta_e$, $\beta_r$ and $\beta_m$ can be obtained from simulations, as the true PSF parameters are known.
We calculate them by measuring the slope of $\gamma_{\rm{obs}}$ as a function of the $\delta e_{p}^{i} r_{p}^2$ and $e_{p}^{i} \delta r_{p}^2$ for $\beta_e^i$ and $\beta_r^i$, respectively.
These coefficients should not differ much between the DES data and simulations, as the estimators should respond in the same way to PSF errors for similar galaxy samples.
Coefficient $\beta_m$ can be obtained from simulations by measuring the slope of multiplicative shear bias as a function of $\delta r$.

We aim for each of these terms to have only a small contribution to the shear $C_{\ell}$, such that the systematic error is smaller than roughly half of the statistical error.
For $N_{\ell}=15$ data points chosen in this analysis, this corresponds to a systematic contribution by less than $0.5/\sqrt{N_{\ell}} \approx 0.15$ of the statistical error to each $C_{\ell}$ vector element;
for example $\alpha^2 C_{\ell}^{ e_{p}^{i} r_{p}^2 } < 0.5 \sigma[ C_{\ell}^{\gamma} ] / \sqrt{N_{\ell}}$.
Note that to achieve a 50\% shift in contours relative to the statistical uncertainty for our $C_{\ell}$ vector, all its elements would have to shift consistently by 15\% in the same direction, if covariance is neglected.
We consider this requirement with respect to a measurement that does not include the marginalisation of systematic error contributions from shear calibration and $n(z)$ uncertainty.
As this marginalisation significantly increases the constraints, our requirement can be considered as conservative.
Our requirement ignores the cross correlations between the elements of the $C_\ell$ vector.
As the power spectrum is fairly independent (see Section \ref{sec:covmat}) and dominated by the shape noise, we do not expect our requirement calculation to be significantly affected by this simplification.

The cross power spectrum between the measured shear $\gamma_{\rm{obs}}$ and each of the additive terms will scale linearly with the coefficient, for example \
$C_{\ell}^{ \gamma \times e_{p}^{i} r_{p}^2 } = \alpha C_{\ell}^{ e_{p}^{i} r_{p}^2 }$.
We use this relation to check the level of systematic contribution against the requirement stated above, by comparing the cross power spectra systematic uncertainty divided by a corresponding coefficient.
For example, for the contribution of the deconvolution error, we have
\begin{equation}
C_{\ell}^{ \gamma \times e_{p}^{i} r_{p}^2 } < \frac{0.5}{\sqrt{N_{\ell}}} \frac{\sigma[ C_{\ell}^{\gamma} ]}{\alpha}.
\end{equation}
Cross power spectra  $\gamma \times \delta e_{p}^{i} r_{p}^2$ and $\gamma \times e_{p}^{i} \delta r_{p}^2$ should satisfy the same condition, divided by coefficients $\beta_e$ and $\beta_r$, respectively.
It is not possible to estimate the multiplicative contribution from cross power spectra, but we can estimate it directly as $\beta_m^2 C_{\ell}^{\delta r_p}$.
We calculate the requirement on the multiplicative bias by comparing the diagonal of the covariance matrix with the amplitude of the signal
\begin{equation}
     C_{\ell}^{\delta r_p} <  \frac{0.5}{\sqrt{N_{\ell}}} \frac{\sigma[ C_{\ell}^{\gamma} ]}{C_{\ell}^{\gamma}  \beta_m^2}.
\end{equation}
We find $ 0.15 \sigma[ C_{\ell}^{\gamma} ] /C_{\ell}^{\gamma} \approx 0.016$ for the $\ell=200$ using the covariance matrix and the fiducial cosmology power spectrum.
Larger $\ell$ have larger requirement.
We ignore the contribution of the cross-correlations between the systematics in this calculation, as we found that they are typically very small.

We measured the following values of the coefficients:
\begin{align*}
  \alpha^{1}_{\mathrm{DES}}, \ \alpha^{2}_{\mathrm{DES}}   &= -0.0122, \ -0.0105 \pm 0.0003 \\
  \alpha^{1}_{\mathrm{UFig}}, \ \alpha^{2}_{\mathrm{UFig}}  &= +0.0001, \ -0.0013 \pm 0.0003 \\
  \beta^{1}_{e}, \ \beta^{2}_{e} &= -0.0646, \ -0.0673  \pm 0.0007 \\
  \beta^{1}_{r}, \ \beta^{2}_{r} &= -0.109, \  -0.040  \pm 0.009 \\
  \beta_m^{1}, \ \beta_m^{2}     &= +0.55, \ +0.39  \pm 0.33. \\
\end{align*}
The leakage parameters $\alpha^{i}$ were calculated from the DES data and the UFig simulations separately.
Remaining coefficients were calculated from the simulations.
The measurement of $\beta_m$ is limited by the number of simulations of the fiducial model, which in our case was \nsurveys.
No significant asymmetry in these coefficients between the two ellipticity components is found, except for $\beta^{i}_{r}$.
In the subsequent calculations we use the higher value out of the two components.
For $\beta_m$ we use the upper limit $| \beta_m | <1$.

\subsubsection{2-pt statistics}
\label{sec:2pt_statistics}

\begin{figure*}[t!]
\includegraphics[width=0.7\linewidth]{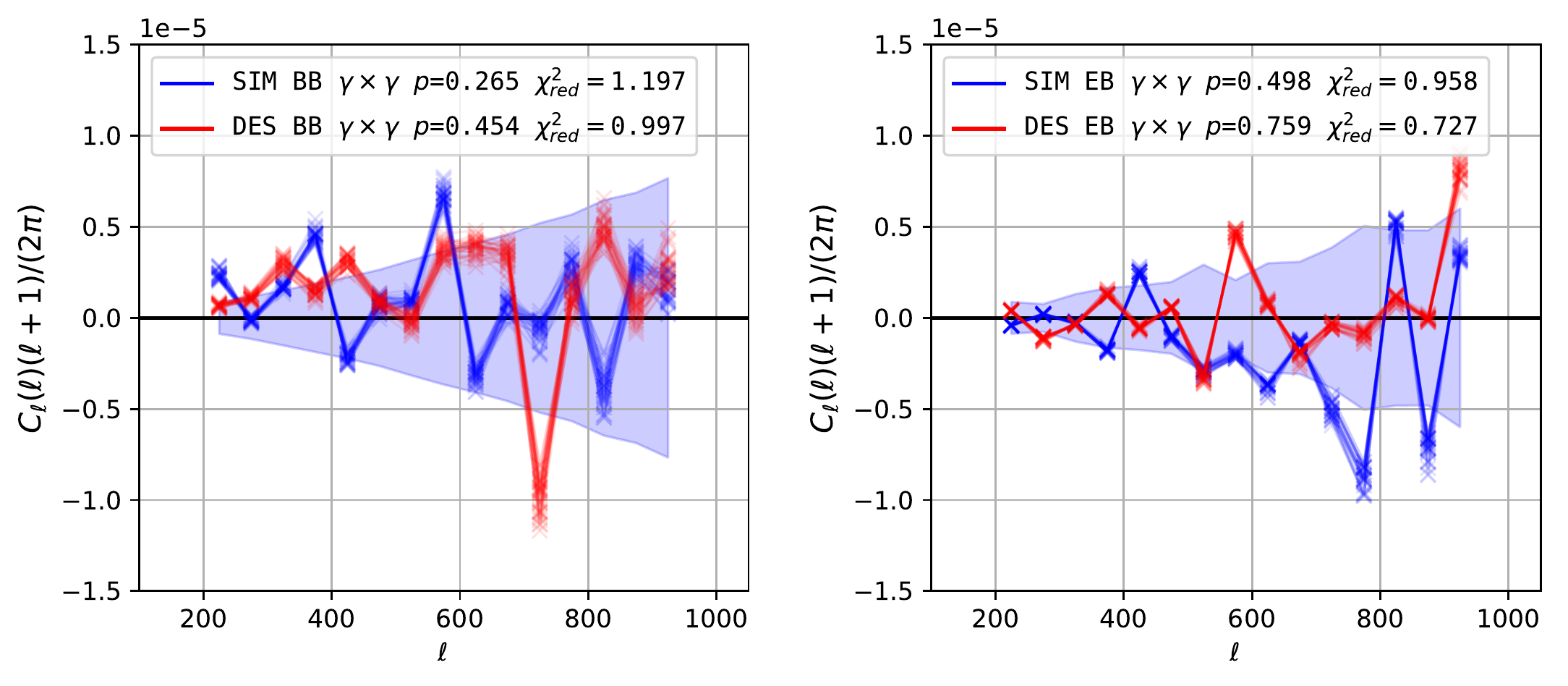}
\caption{
The shear B-mode auto power spectrum (left panel) and BE cross power (left panel).
The red lines correspond to the DES shear catalogues calibrated with \nsurveys\ different calibration factors from the ABC posterior.
The blue lines correspond to fiducial UFig simulation, also calibrated with these \nsurveys\ factors.
The spread between these data points corresponds to the systematic uncertainty.
The shaded region corresponds to $1\sigma$ statistical uncertainty on the measured quantity.
For the BB spectrum, these values are taken as the square root of the diagonal of the covariance matrix (see Section \ref{sec:covmat}).
The uncertainty on the EB spectrum was measured from \nsurveys\ simulations of the fiducial survey, with different random seeds.
}
\label{fig:shear_bmode}
\end{figure*}

We calculated the shear power spectrum B-modes, as well as cross-power spectra between shear and PSF ellipticity and size, and residual of these quantities.
Figure \ref{fig:shear_bmode} shows the shear B-mode auto power spectrum (BB) and cross power spectrum between the E- and B-modes (EB).
Red and blue lines show the measurement for DES and fiducial UFig survey, respectively, for \nsurveys\ calibration parameters found from ABC posterior.
The shaded regions correspond to the statistical uncertainty on the measurement.
The B-mode of the DES data is found to have reduced $\chi^2=0.997$, which corresponds to the $p$-value of $p=0.454$, as calculated using the full B-mode covariance matrix (see Section \ref{sec:covmat}).
The EB cross correlation is also not statistically significant.

The cross spectra of shear E-mode and PSF parameters are shown in Figure~\ref{fig:shear_cross_PSF}.
Again, red and blue lines correspond to the DES and UFig data, respectively.
Blue shaded regions signify the $1\sigma$ uncertainty, measured from random seed realisations of the fiducial UFig survey.
The grey lines correspond to the requirement described in Section~\ref{sec:systematics_model}.
Note that in order to achieve a 0.5$\sigma$ shift in the contours, all the points would have to consistently exceed the requirement in the same direction.
This requirement considers only the statistical error and does not include the systematics, which makes it conservative.

There is a slight correlation of the shear with PSF ellipticity $e_p r_p$ E-mode, with $\chi^2_{\mathrm{red}}=1.79$.
This is consistent with the PSF leakage calculated at the 1-pt level.
We also notice a significant cross power spectrum between the shear E-mode and PSF size residual B-mode, but that seems to be dominated by one outlier point at $\ell \sim 900$.
This trend is not visible for equivalent EE cross power spectrum.
The reduced $\chi^2$ for other cross power spectra are generally close to  $\chi^2_{\mathrm{red}}=1$.
It is important to note that the distribution of the cross power spectra is not expected to be Gaussian, and therefore the reduced $\chi^2$ may not be the best way to quantify the systematic significance for this problem.
We leave this investigation to future work.
Nevertheless, none of these cross correlations consistently exceed the given requirements calculated with coefficients calculated in Section~\ref{sec:systematics_model} in our considered $\ell$ range.
This suggests that the PSF was removed well enough for our requirements.
The multiplicative error resulting from the error in the PSF size measurement will depend on the coefficient $\beta_m$, calculated in Section~\ref{sec:systematics_model} and the PSF size error power spectrum, calculated in Section~\ref{sec:systematics_maps}, so that $C_{\ell}^{\gamma, \mathrm{obs}} = C_{\ell}^{\gamma, \mathrm{true}} (1+ \beta_m^2 C_{\ell}^{\delta r_p})$.
We measured $C_{\ell}^{\delta r_p}<2 \e{-4}$ and $| \beta_m | < 1$, and therefore we expect the multiplicative error stemming from the PSF size error to be smaller than the requirement.

These requirements, however, were exceeded for $\ell$ outside considered range.
For low $\ell$, the leakage contribution started to exceed the requirement.
For high $\ell$, the PSF ellipticity residual increased greatly which lead to $\gamma \times (e_{p}-e_{s}) r_p^2 $ exceeding the requirements.
We therefore decided to limit the $\ell$ range in our analysis to $\ell \in [200, 950]$.

The EE power spectrum measured from the DES source galaxy sample is shown on the left panel in Figure~\ref{fig:posterior_cls_nz}.
The black line corresponds to the fiducial survey found by the ABC scheme (see Section~\ref{sec:abc-fits}).
The error-bars show the standard deviation estimate, calculated by taking the square root of the diagonal of the covariance matrix.
The full covariance matrix of $C_{\ell}$ is shown in Section~\ref{sec:covmat}.
The red lines correspond to the $C_{\ell}$ calculated from the DES sample, but calibrated using \nsurveys\ different parameter sets, which were obtained from different ABC posterior points.
The size of the source galaxies catalogue was varying slightly within the \nsurveys\ sets, as each time we removed galaxies with large ellipticities $|e|>1$.
The spread between the \nsurveys\ $C_{\ell}$ measurements represent the systematic uncertainty in our calibration scheme.
The $C_{\ell}$ corresponding to the best fit cosmological model for the fiducial survey is shown in grey line.
The reduced $\chi^2$ of this fit is  $\chi^2=0.813$.

\subsection{Redshift measurement}
\label{sec:nz-measurement}
In this analysis we use a single redshift bin, without tomography.
We apply the method given in \cite{Herbel2017redshift} to infer the redshift distribution of our lensing sample.
The key idea of this method is to make realistic simulations of the observed data and apply exactly the same image processing pipeline to the observed and simulated data.
This way the detection method and the cuts are the same, which allows us to calculate the $n(z)$ distribution by simply taking the true redshift of all simulated galaxies in the sample.
The ABC uses distance measures based on properties of galaxies in the image data, as well as the distribution of redshifts in the VVDS spectroscopic sample, to obtain the posterior distribution on the input parameters, and consequently corresponding $n(z)$.
Note that this method can only be use to calculate the $n(z)$ distribution of a sample of galaxies, not redshifts of individual galaxies.

\begin{figure*}[t!]
\includegraphics[width=0.7\linewidth]{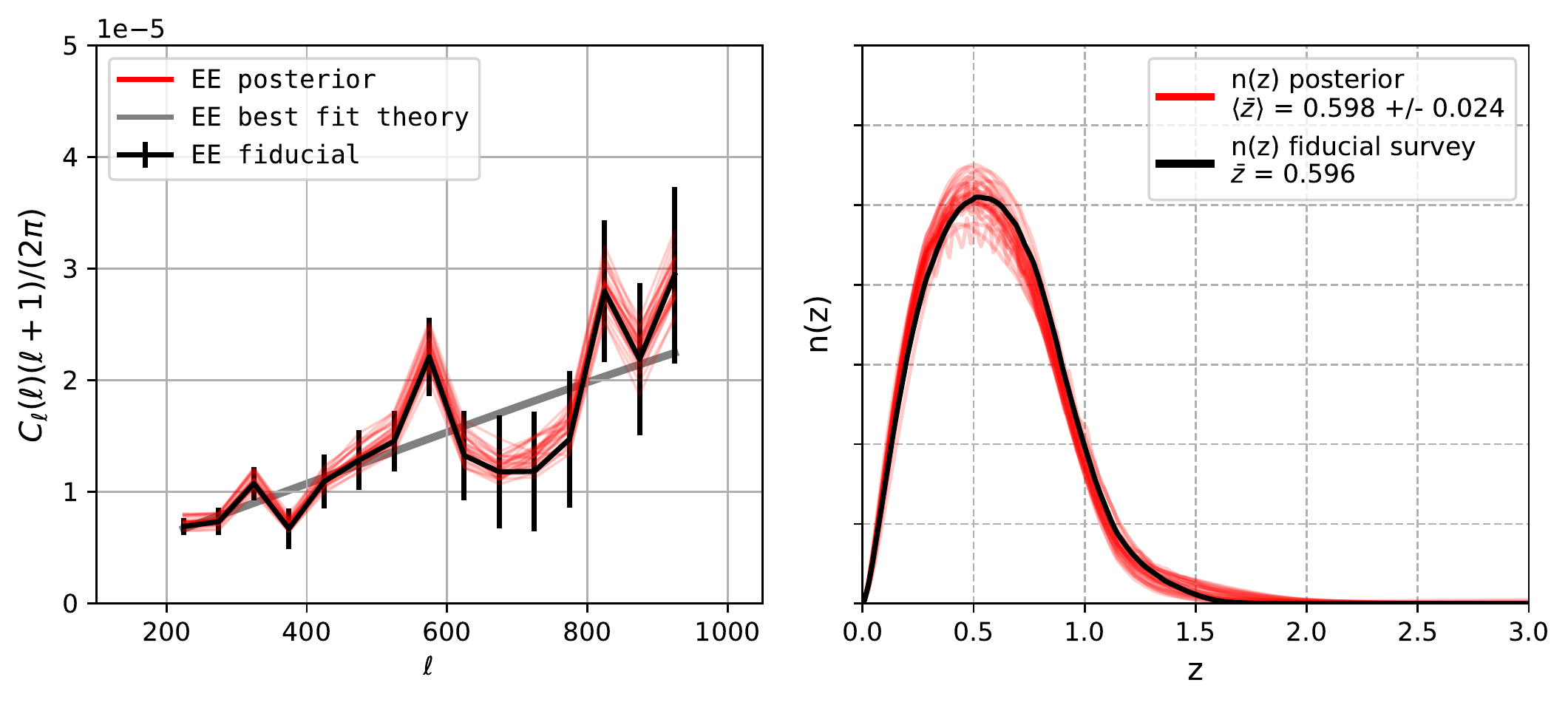}
\caption{
Joint n(z) and shear $C_\ell$ measured from the DES Y1 data set.
Multiple red lines correspond to points from the ABC posterior.
The black line shows the measurement using the calibration parameters calculated from the fiducial survey.
The error-bars correspond to $1\sigma$ errors and were taken from the diagonal of the covariance matrix (see Section~\ref{sec:covmat}).
The best fit theory model for the fiducial survey is shown with the grey line.
}
\label{fig:posterior_cls_nz}
\end{figure*}

In this work, we obtain a posterior set of likely $n(z)$ curves for our lensing sample from the ABC posterior described in section \ref{sec:abc-fits}.
The redshifts of the galaxies in the simulated lensing samples are then used to construct a family of \nsurveys\ $n(z)$ curves.
This result is shown in the right panel of Figure \ref{fig:posterior_cls_nz}.
We compute an average $\langle \bar z \rangle$ of mean redshifts $\bar z_i$ of each curve.
Using that ensemble we compute the $\langle \bar z \rangle = 0.598 \pm 0.024$.
The mean $n(z)$ for the fiducial survey was $\bar z = 0.596$.

A tomographic approach could be straightforwardly accommodated in the MCCL framework.
It would nevertheless require more development of the pipeline, which we leave for future work on the following DES data releases.

\subsection{Robustness to model extensions (loop 3.2)}
\label{sec:loop32}
The robustness of the measurement can be tested against extensions to the forward model (loop 3.2).
If the measurement does not significantly respond to the extension, there is no need to incorporate it into the model.
Here we performed one such extension in this work for the fiducial simulation.
We allowed scatter in the S\'ersic index of blue and red galaxies.
The S\'ersic index was drawn from a truncated normal distrubtion in the limits between 0.5 and 5.
The mean, before truncation, was set to the original value obtained with ABC, and the standard deviation was set to 0.5.
We ran a new simulation with this model using the fiducial parameter set and calculated calibration parameters and $n(z)$ for the extended model the same way as before.
We obtained $\alpha_1=0.731$, $\eta=0.783$, $\bar z=0.594$.
These parameters are close to the ones calculated from the base model and within the uncertainty calculated from \nsurveys\ surveys.
This indicates that the measurement is robust to this extension and we do not incorporate it into the base model.

\subsection{Discussion}
\label{sec:remaining_issues}

The mean shear for the DES and UFig catalogue calculated with fiducial calibration factor is
\begin{align*}
\langle \gamma_1^{\rm{DES}}\rangle, \ \langle  \gamma_2^{\rm{DES}} \rangle &= -0.00203, \ 0.00077 \pm 0.00007 \\
\langle \gamma_1^{\rm{SIM}}\rangle, \ \langle  \gamma_2^{\rm{SIM}} \rangle &= -0.00049, \ 0.00053 \pm 0.00007. \\
\end{align*}
The dispersion in the mean shear from cosmic variance is expected on the level of $\sigma[\langle  \gamma \rangle] \approx 2 \cdot 10^{-4}$.
The mean shear in DES catalogue is significantly higher than expected from the cosmic variance.
For simulations, the mean shear is consistent with cosmic variance.
The mean shear, however, does not influence the power spectrum analysis, as we found that it affects only ${\ell<100}$, and is subtracted before calculating the power spectrum.
More details about 1-pt statistics can be found in Appendix~\ref{appendix:shear_1pt}.
Nevertheless, it can be a sign of remaining systematics.
We looked for the source of this mean shear by examining its dependence of various effects that can cause systematic ellipticity shifts.
We considered 17 possible variables, including brightness, colours, PSF parameters, distance to bright stars, and position in the footprint.
We did not find any significant trends that were not present in the simulations.
Given that the mean shear seems to be independent of the systematic variables, we conclude that it should not have a significant impact on the 2-pt measurements.

\section{Cosmological constraints}
\label{sec:cosmo_constraints}

We measure the cosmological parameters from the non-tomographic shear power spectra.
We focus on the flat $\Lambda$CDM model and vary 5 cosmological parameters $\boldsymbol{\theta} = \{h, \Omega_{\mathrm{m}}, \Omega_{\mathrm{b}}, n_{\mathrm{s}}, \sigma_{8}\}$, where $h$ is the dimensionless Hubble parameter, $\Omega_{\mathrm{m}}$ is the fractional matter density today, $\Omega_{\mathrm{b}}$ is the fractional baryon density today, $n_{\mathrm{s}}$ denotes the scalar spectral index and $\sigma_{8}$ is the r.m.s. of linear matter fluctuations in spheres of comoving radius $8 \,h^{-1}$ Mpc.

\subsection{Theory prediction with PyCosmo}

To compute theoretical predictions for cosmic shear power spectra, we follow ~\cite{Nicola:2016, Nicola:2017} and use the Limber approximation \cite{Limber:1953, Kaiser:1992, Kaiser:1998}, which is a valid approximation for large multipoles, typically $\ell > \mathcal{O}(10)$, and broad redshift bins \cite{Peacock:1999}.
The expression for the spherical harmonic power spectrum $C_{\ell}^{\gamma\gamma}$ at angular multipole $\ell$ can be found in \citep{Nicola:2016}.
We compute all theoretical predictions using PyCosmo \citep{Refregier2017pycosmo} and use the transfer function derived by \cite{Eisenstein:1998} to calculate the linear matter power spectrum.
To compute the non-linear matter power spectrum we use the $\textsc{HMcode}$ fitting function \cite{Mead:2015, Mead:2016}.
To account for the effects of the survey mask, we multiply the predicted $C_{\ell}$ with the \textsc{PolSpice} kernels, which were outputted during the $C_{\ell}$ calculation from the survey data (\textsc{PolSpice} command argument \verb+kernelsfileout+).
The power spectrum was then binned into 15 linearly spaced bins, from $\ell_{\rm{min}}=200$ to $\ell_{\rm{max}}=950$.
The choice of scales was informed by requirements on systematic errors, described in Section~\ref{sec:systematics_model}.

\subsection{Intrinsic Alignments}
\label{sec:IA}

To model the intrinsic alignments, we implement the `non-linear linear alignment model' \cite{HirataSeljak2004intrinsic}, which was used in \citep{des2015sv,Hildebrandt2016kids,Heymans2013tomographic}.
It consists of two contributions, from intrinsic-intrinsic ($II$) and shear-intrinsic ($GI$) shape correlations, so that the measured $C_{\ell}^{\rm{obs}}$ is
\begin{equation}
    C_{\ell}^{\rm{obs}} = C_{\ell}^{\gamma \gamma} + A_{\mathrm{IA}}^{2}C_{\ell}^{II} + A_{\mathrm{IA}}C_{\ell}^{GI},
\end{equation}
where $A_{\mathrm{IA}}$ is the intrinsic alignment amplitude parameter.
The impact of $A_{\mathrm{IA}}$ is almost perfectly degenerate with the $\sigma^2_8$ for a non-tomographic measurement in the $C_{\ell}$ range considered.
That is why, in this work, we put constraints on a combination of $\sigma_8$ and $A_{\mathrm{IA}}$, and report the product $\sigma_8 D_{\mathrm{IA}}$, where $D_{\mathrm{IA}}$ is a scaling factor dependent on the strength of intrinsic alignment $A_{\mathrm{IA}}$, as described below.

To do this, we first calculate the ratio between the intrinsic alignment and shear power spectra for our fiducial cosmological model parameters and $A_{\mathrm{IA}}=1$.
We found that it is a constant fraction to a good approximation: $C_{\ell}^{II}=f_1 C_{\ell}^{\gamma \gamma}$ and  $C_{\ell}^{GI}=f_2 C_{\ell}^{\gamma \gamma}$, where $f_1=0.019$ and $f_2=-0.117$.
As $\sigma^2_8$ is proportional to $C_{\ell}^{\gamma \gamma}$, $\sigma_8$ will depend on deviation from $A_{\mathrm{IA}}=1$ in the following way
\begin{equation}
\sigma_8(A_{\mathrm{IA}}=1) \approx \sigma_8(A_{\mathrm{IA}}) \left[ \frac{1+ f_1 A_{\mathrm{IA}}^2 + f_2 A_{\mathrm{IA}}}{1+f_1+f_2} \right]^{1/2}
\end{equation}
and at the fiducial cosmology can be further linearised to:
\begin{equation}
\label{eqn:IA_scaling}
\sigma_8(A_{\mathrm{IA}}=1) \approx \sigma_8(A_{\mathrm{IA}}) \cdot D_{\mathrm{IA}}
\end{equation}
where \defDIA\ is the parameter that scales $\sigma_8$ according to the intrinsic alignment amplitude.

\subsection{Baryonic corrections}
\label{sec:baryonic_corrections}

For the baryonic corrections we use prescription in \citep{Mead:2015}, who parametrises the effects of baryons on the non-linear power spectrum.
Specifically, we follow the implementation of \citep{Hildebrandt2016kids}, who uses a flat prior in the range $B_{\rm{baryon}} \in [2,4]$ of \citep{Mead:2015}.
This range corresponds to feasible range associated with different baryonic feedback models, as seen with hydrodynamic simulations.
The dark matter only case corresponds to $B_{\rm{baryon}}=3.13$.
The second baryon correction parameter $\eta_0$ in \citep{Mead:2015} is set by the equation 30 in that paper.

\subsection{Covariance matrix estimation with \textsc{L-Picola}}
\label{sec:covmat}

\begin{figure}[t!]
\includegraphics[width=1\linewidth]{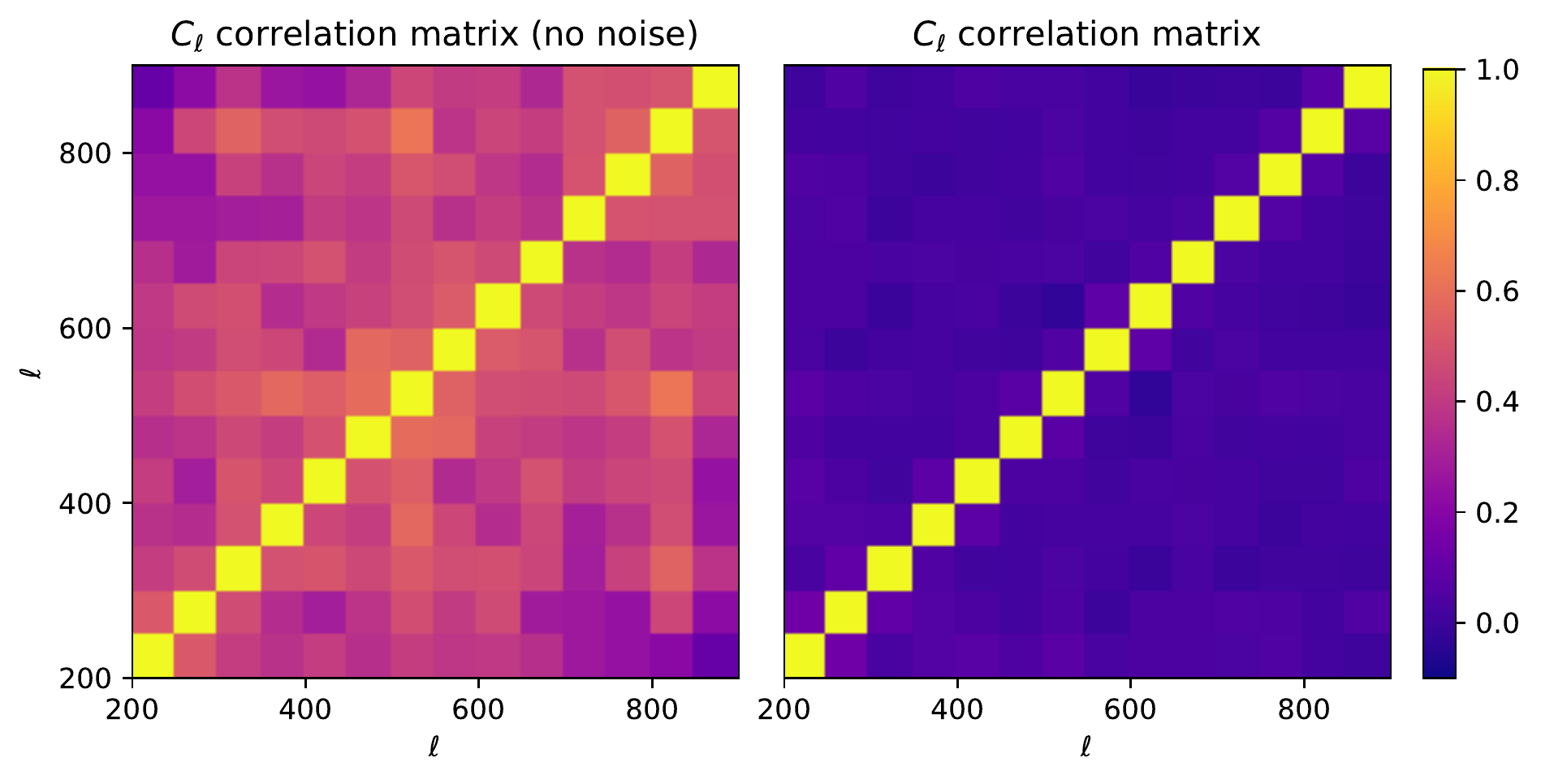}
\caption{Correlation coefficient matrix of the power spectrum for angular scales $10^2 < \ell < 10^3$ using the $n(z)$ for the fiducial survey (see Section \ref{sec:nz-measurement}) and a binning of $\Delta \ell = 50$ using 1000 realisations including shape noise in the right plot and using 100 realisations without shape noise in the left plot.}
\label{fig:correlation}
\end{figure}

\begin{figure*}[t!]
\includegraphics[width=0.8\linewidth]{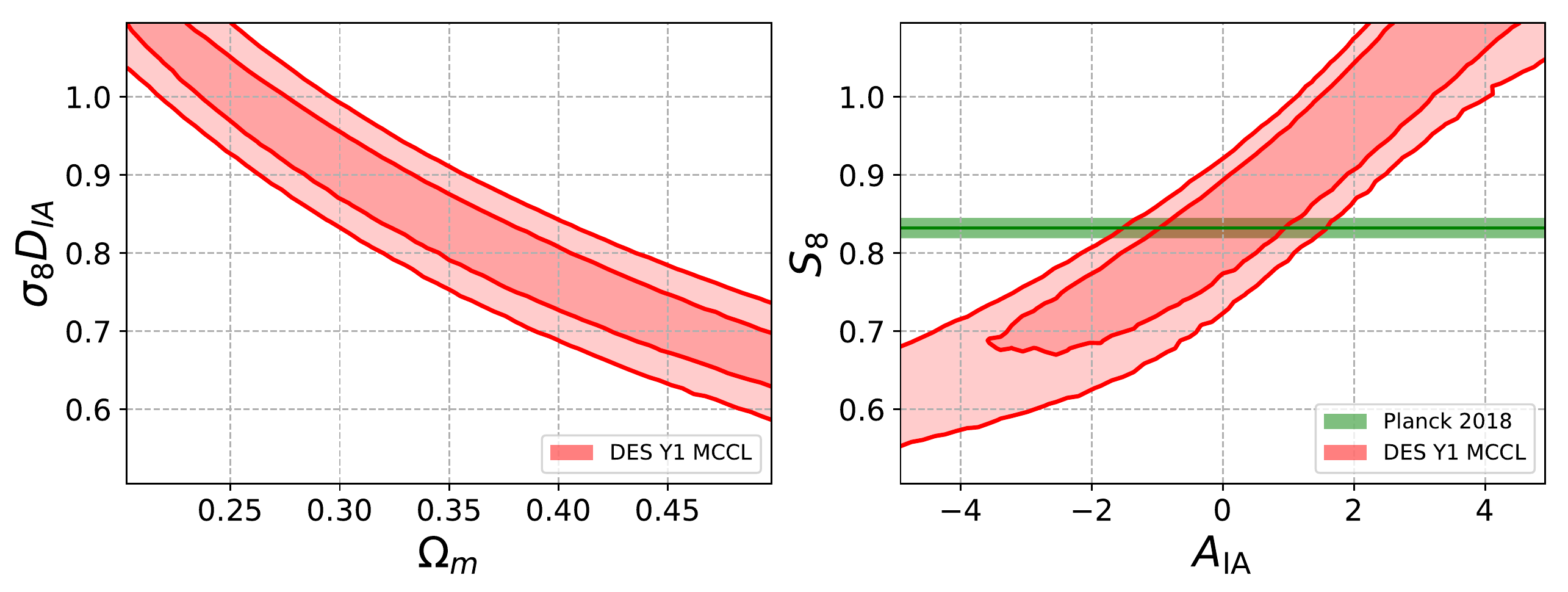}
\caption{Cosmological constraints on $\Omega_m$ and $\sigma_8 D_{\mathrm{IA}}$ marginalised over $n(z)$ and shear uncertainties (left panel).
The parameter \defDIA\ depends on the amplitude of intrinsic alignments (see Section~\ref{sec:IA}).
The lines show the 68\% and 95\% confidence intervals.
The right panel shows the $S_8$ parameter constraint as a function of the intrinsic alignment amplitude.
The green bar corresponds to the Planck 2018 measurement of $S_8$ (TT,TE,EE + lowE + lensing) \citep{Planck2018parameters}.
}
\label{fig:constratints}
\end{figure*}

We compute the covariance matrix used for the likelihood analysis by following the method described in \cite{Sgier2018}.
The method used here avoids the survey geometry-related effects described in \citep{Troxel2018geometry}.
The matter density field is simulated using the fast approximate $N$-Body code \textsc{L-Picola} \cite{Howlett2015} with the number of particles $N_\text{part}=1024$ and the mesh density set to $N_\text{mesh}=2048$ per side of the simulation volume.
The all-sky past-lightcone is constructed by fixing the observer at the centre of the simulation volume and by slicing the volume between $z=0.0$ and the final redshift of $z=1.5$ with no gaps into comoving concentric spherical shells of thickness $\Delta \chi_b = \chi(z_b + \Delta z) - \chi (z_b)$ with a redshift-shell thickness of $\Delta z = 0.01$, whereas $z_b$ is the redshift of the particles within the shell of index $b$.
We do not replicate the simulation volume for this construction, since this could potentially lead to statistical artefacts in the power spectra and therefore in the covariance matrix. Furthermore, in this way we ensure that super-survey modes are correctly captured.
In order to obtain accurate results for the spherical harmonic power spectra up to angular scales of $\ell \sim 10^3$, we nest three simulation volumes with different box-sizes $L_1 = 700 \, \text{Mpc}/h$, $L_2 = 4.2 \, \text{Gpc}/h$ and $L_3 = 6.3 \, \text{Gpc}/h$. This is done by first constructing the lightcone in the smallest simulation volume from redshift $0.0$ to its edge at $0.1$ and then continuing the construction in the larger volume from $0.1$ to $0.8$ using $L_2$ and from $0.8$ to $1.5$ using $L_3$.

We run 10 \textsc{L-Picola} realisations with different seeds (running three simulations with box-sizes $L_1$, $L_2$ and $L_3$ per realisation) and applied the above described pipeline to calculate 10 full-sky convergence maps using the $n(z)$ distribution obtained using the method described in section \ref{sec:nz-measurement}.
Each map is processed with the DES Y1 survey mask resulting in 10 non-overlapping patches on the sphere.
We then add 10 different shape noise realisations to each patch and compute the cosmic shear power spectra (see Appendix~\ref{appendix:power_spectrum_calculation}).
The shape noise was created by rotating the galaxy ellipticity by a random angle.
The covariance matrix used for the analysis it then computed using 1000 realisations (10 maps $\times$ 10 patches $\times$ 10 noise realisations) of the cosmic shear power spectra for the same binning as for the measurement.
The resulting correlation matrix for the shear power spectrum is shown in Figure~\ref{fig:correlation}.
The left panel shows the $C_{\ell}$ correlation matrix when no shape noise is added, while the right panel includes the shape noise, which was used in the likelihood.
We notice that the covariance matrix is dominated by the shape noise contribution.

\subsection{Blinding}
\label{sec:blinding}

Throughout the analysis we follow the blinding scheme applied to the EE shear power spectrum, described in Appendix \ref{appendix:blinding}.
We define the following set of quality conditions to be met by our measurement before unblinding:
(Q1) properties of the DES galaxy population have to lie within the space covered by the simulations,
(Q2) for the fiducial simulation, the input cosmology should be accurately recovered,
(Q3) the impact of the possible discrepancies in systematic maps can not be larger than the statistical errors on the measurement,
(Q4) small shear B-mode and small cross power spectra between shear and PSF, and
(Q5) analysis versions that include well-motivated extensions to the model should not cause significant difference in final result.
We review the satisfaction of these conditions in Appendix~\ref{appendix:blinding} and conclude that these conditions were met.
Additional actions taken after the unblinding are documented (Appendix \ref{appendix:blinding}).

\subsection{Constraints}

\renewcommand{\arraystretch}{2.0}
\begin{table*}
\setlength{\tabcolsep}{5pt}
\footnotesize
    \centering
    \begin{tabular}{ p{0.65\linewidth} | c p{0.25\linewidth} }
        \hline
        analysis variant                     &    $S_8 D_{\mathrm{IA}}=\sigma_8(\Omega_m/0.3)^\alpha D_{\mathrm{IA}}$ \\
        \hline

MCCL non-tomo with marginalised systematics                                                          & $0.895^{+0.054}_{-0.039}$ \\
MCCL non-tomo with statistical uncertainty only                                                      & $0.881^{+0.050}_{-0.033}$ \\
MCCL non-tomo with marginalised systematics, $\alpha$=0.6                                            & $0.907^{+0.047}_{-0.040}$ \\
MCCL non-tomo, \textsc{Halofit}, no baryons                                                          & $0.880^{+0.047}_{-0.038}$ \\
\textsc{Im3shape} non-tomo, $C_{\ell}$ analysis                                                      & $0.857^{+0.063}_{-0.048}$ \\
\textsc{Im3shape} non-tomo, \textsc{Halofit}, no baryons, $C_{\ell}$ analysis                        & $0.846^{+0.054}_{-0.045}$ \\
DES Y1 cosmic shear fiducial (\textsc{Metacalibration}), tomographic (Troxel et al. 2017)            & $0.782^{+0.027}_{-0.027}$ \\
DES Y1 \textsc{Im3shape}, tomographic (Troxel et al. 2017)                                           & $0.799^{+0.048}_{-0.045}$ \\

    \hline
    \end{tabular}
    \caption{Comparison of $S_8 D_{\mathrm{IA}}$ measurements from the non-tomographic MCCL analysis with the tomographic DES Y1 cosmic shear \citep{Troxel2017cosmological}, which is additionally marginalised over the uncertainy in the intrinsic alignment modelling.
    Parameter \defDIA\ controls the strengths of intrinsic alignment, so that $D_{\mathrm{IA}}=1$ for $A_{\mathrm{IA}}=1$.
    The $\alpha$ parameter in the $S_8$ definition was set to $\alpha=0.5$, similarly to the previous DES Y1 analysis \citep{Troxel2017cosmological}, unless otherwise stated.
    We found the best-fitting $\alpha$ parameter in the $S_8$ definition to be $\alpha=0.6$.
    See Section~\ref{sec:comparison_des} for details about the $C_{\ell}$ analysis with \textsc{Im3shape}.
     }
    \label{tab:cosmo_params_comparison}
\end{table*}

We obtain the posterior distribution on cosmological parameters for each of the \nsurveys\ $n(z)$ and $C_{\ell}$ pairs.
We create the final constraint with $n(z)$ and shear calibration uncertainty marginalised by adding together the \nsurveys\ posterior probabilities.

To compute cosmological parameter constraints, we assume the cosmic shear likelihood to be Gaussian, with log-likelihood of the following form
\small
\begin{align}
\mathscr{L}(D \vert \theta) = &
{
\frac{1}{|2 \pi \Sigma |^{\frac{1}{2}}}
-\frac{1}{2}
\left( C^{\mathrm{obs}}_{\ell} - C^{\mathrm{the}}_{\ell} \right)^{\mathrm{T}}
\mathrm{\Sigma}^{-1}
\left( C^{\mathrm{obs}}_{\ell} - C^{\mathrm{the}}_{\ell} \right) }
 \end{align}
\normalsize
where $C^{\mathrm{obs}}_{\ell}$ and $C^{\mathrm{the}}_{\ell}$ are the power spectra from observations and theory, respectively, $\Sigma$ denotes the covariance matrix, and $d$ is the size of the data vector.

We assume flat priors on the cosmological parameters:
$h \in [0.5,  1.0]$,
$\Omega_m \in [0.2,  0.5]$,
$\Omega_b \in [0.01, 0.09]$,
$n \in [0.75, 1.25]$,
$\sigma_8 \in [0.5,  1.1]$,
$B_{\rm{baryon}} \in [2.0,  4.0]$.
The intrinsic alignment parameter was fixed to $A_{\mathrm{IA}}=1$ for the main result, or sampled inside the uniform prior $A_{\mathrm{IA}}\in[-5,5]$.

To obtain the posterior distribution, we evaluate 200,000 likelihoods in 6D parameter space at points chosen from a Sobol sequence \citep{Sobol1967sequence}.
A Sobol sequence is a set of points in a high-dimensional space designed such that the projections of the space are sampled as regularly as possible.
Sobol sequence integration is an effective technique to approximate multidimensional integrals \citep{Kucherenko2015exploring}.
The Sobol sequence was continued for each of the \nsurveys\ $C_{\ell}$ and $n(z)$ pairs, such that likelihoods for each pair were evaluated at different points in this space.
We use this technique instead of Monte Carlo Markov Chain (MCMC), as the total turnaround time of this calculation, including queuing time, with single-core jobs is faster and more stable than the parallelised MCMC, despite larger number of likelihood evaluations.
To create a 2D posterior for a cosmological parameter pair, we calculate a 2D histogram of all the Sobol sequence points, weighted by their probability.
That histogram is then normalised and the confidence intervals are calculated.
The posteriors from 30 surveys capture the uncertainty from $n(z)$ and shear calibration, and to marginalise this uncertainy we add the 30 PDFs.
This corresponds to discretising the integral
\begin{equation}
    p(\theta_{1}) = \int d\theta_{2} p(\theta_{1}|\theta_{2}) p(\theta_{2}) \approx \sum_{\theta_2^{i} \sim p(\theta_2)} p(\theta_{1}|\theta_{2}^{i})
\end{equation}
where $\theta_{1}$ are the parameters of interest and $\theta_{2}$ are the nuisance parameters, and $\theta_2^{i} \sim p(\theta_2)$ is a set of samples from $p(\theta_2)$.
Note that individual distributions $p(\theta_{1}|\theta_{2}^{i})$ are not normalised.

In this non-tomographic, cosmic shear only analysis we can only effectively constrain the combination of $\Omega_m$, $\sigma_8$, and $A_{\mathrm{IA}}$.
Other parameters remain unconstrained.
The left panel on Figure~\ref{fig:constratints} shows the constraints in the $\Omega_m-\sigma_8 D_{\mathrm{IA}}$ plane with marginalised uncertainty on $n(z)$ and shear calibration.
The lines represent 68\% and 95\% confidence intervals.
The shape of the contour follows a degeneracy characteristic for $\Omega_m-\sigma_8$.
We calculate the constraint on the combination of these parameters $S_8 D_{\mathrm{IA}} =\sigma_8(\Omega_m/0.3)^{0.5} D_{\mathrm{IA}}$.
For this fiducial configuration, we find ${S_8 D_{\mathrm{IA}}=0.895^{+0.054}_{-0.039}}$.
The right panel on Figure~\ref{fig:constratints} shows the dependence of the $S_8$ constraint on the intrinsic alignment amplitude $A_{\mathrm{IA}}$.
There is a clear degeneracy between those parameters.
The constraint from the Planck survey \citep{Planck2018parameters} is shown with the green bar (TT,TE,EE + lowE + lensing).

We calculate the results for other analysis variants and summarise them in Table~\ref{tab:cosmo_params_comparison}.
When the systematic uncertainty is ignored, we find
${S_8D_{\mathrm{IA}}=0.881^{+0.050}_{-0.033}}$
for the fiducial survey.
This indicates that the contribution of the systematic uncertainty to the error budget is on the level of 60\% of the statistical uncertainty.
The main source of systematic uncertainty is the shear calibration, but it does not dominate it completely.
We calculated the standard deviation $\sigma[C_{\ell}^{\mathrm{the}}]$ of theory power spectrum predicted from the same cosmological parameter set using \nsurveys\ $n(z)$,  and the standard deviation of the measurement $\sigma[C_{\ell}^{\mathrm{obs}}]$ when \nsurveys\ calibration parameters are applied.
We find that ${\sigma[C_{\ell}^{\mathrm{the}}] /  \sigma[C_{\ell}^{\mathrm{obs}} ]\approx0.6} $ for large scales, and decreases to $\approx 0.05$ for small scales.

We give the main result for the ${\alpha=0.5}$ in the $S_8=\sigma_8 \left( \Omega_m/0.3 \right)^{\alpha}$ definition in order to easily compare with other measurements.
We find, however, that the best-fitting value of the $\alpha$-parameter is ${\alpha=0.6}$ and gives slightly narrower constraint,
${S_8D_{\mathrm{IA}}=\sigma_8(\Omega_m/0.3)^{0.6}D_{\mathrm{IA}}=0.907^{+0.047}_{-0.040}}$.

\subsection{Comparison with previous DES measurements}
\label{sec:comparison_des}

The width of the contours is larger in our analysis than in main DES Y1 \citep{Troxel2017cosmological}, mainly due to different choices of the data vector.
Firstly, we used a single tomographic bin, as opposed to four bins used in \citep{Troxel2017cosmological}.
This results in weaker constraining power in our analysis, roughly by a factor of two.
While no non-tomographic constraints were measured in the main DES Y1 analysis, the Science Verification \citep{des2015sv} analysis compared the fiducial 3-bin tomographic and non-tomographic measurements.
The constraining power was found to decrease roughly by a factor of two.
We find a similar error scaling between our non-tomographic MCCL analysis and the 4-bin tomography in DES Y1.
Secondly, the choice of scale cuts were slightly different: the DES Y1 used scales $<7$ arcmin for $\xi_{+}$, which corresponds to $\ell>1500$.
The range of $\xi_{-}$ was restricted to large angular scales.
The minimum scale was different in each redshift bin.
We restricted our analysis to $\ell \in [200, 950]$.

Detection and measurement of objects in the DES images was a part of the MCCL method.
We used the \band-band only for creating the source galaxy sample, and all five bands ($grizY$) to obtain the ABC posterior on the parameters of the forward model.
Our fiducial source sample contained
15,432,057 galaxies
after applying the cuts described in Appendix~\ref{appendix:galaxy_sample_selection}.
This is similar to $\sim$22 million were selected from the $r$-band only \textsc{Im3shape} catalogue in the main Y1 analysis \citep{Troxel2017cosmological,Zuntz2017catalogues}.
The \textsc{Metacalibration} sample was larger and contained $\sim$34 million objects using joint measurements from three bands.
This results in a higher mean $n(z)$ from 4 tomographic bins with $\langle z \rangle \approx 0.67$ (based on Figure 16 in \citep{Troxel2017cosmological}).
This is higher than the $n(z)$ we obtained, which was $\langle \bar z \rangle = 0.598 \pm 0.024$.
The uncertainty on both mean redshift is close to that calculated for \textsc{Im3shape} by \citep{Hoyle2018redshift}, which ranged from $\sigma(\Delta z) \in (0.11, 0.22)$ among tomographic bins.
The error on the shear calibration is also close; \citep{Zuntz2013im3shape} calculated $\sigma(m)=0.025$, while the spread of the calibration parameter $\eta$ (see Section~\ref{sec:shear_calibration}) in our ABC posterior is $\sigma(\eta)=0.022$.

Finally, we do not include marginalisation over the intrinsic alignment amplitude and present a combination of $S_8 D_{\mathrm{IA}}$.
In the tomographic analysis of DES Y1 by \citep{Troxel2017cosmological}, the uncertainty increased by $\sim 30\%$ due to marginalisation over wide IA prior.
Marginalisation over intrinsic alignment in the non-tomographic analysis would lead to a more significant increase of the uncertainty on the cosmology parameters.
Moreover, the DES analysis by \citep{Troxel2017cosmological} uses \textsc{Halofit}~\citep{Smith2003halofit,Takahashi2012halofit} to model the non-linear power spectrum and removes the dependence on the Baryons by applying scale cuts.
In our analysis, we used the \textsc{HMcode} model by \citep{Mead:2016} and marginalise over a wide prior on Baryonic corrections.

The tomographic DES Y1 analysis by \citep{Troxel2017cosmological} found
${\sigma_8 (\Omega_m/0.3)^{0.5}=0.782^{+0.027}_{-0.027}}$
with the \textsc{Metacalibration} data, and
${\sigma_8 (\Omega_m/0.3)^{0.5}=0.799^{+0.048}_{-0.045}}$
with the \textsc{Im3shape} data.
We investigate the main source of the difference between the MCCL and \textsc{Im3shape} results by comparing the theory modelling, shear calibration, and $n(z)$ measurement, between the pipelines.
We calculate constraints for the MCCL $C_{\ell}$ and $n(z)$ using the \textsc{Halofit}~\citep{Takahashi2012halofit} theory prediction, without marginalisation of the strength of Baryon effects, keeping the rest of the analysis configuration the same as for the main result.
We find $S_8D_{\mathrm{IA}}~=~0.880^{+0.047}_{-0.038}$ for that configuration, which is a small shift at the 0.5$\sigma$ level from the value obtained with \textsc{HMcode} in the fiducial configuration.

We calculate the cosmological constraints from the \textsc{Im3shape} shape catalogue and the corresponding $n(z)$, with the likelihood analysis used for the MCCL result.
To do it, we create a non-tomographic source galaxy sample with a single $n(z)$ bin, by adding four tomographic $n(z)$, weighted by the corresponding effective galaxy densities \citep[see][for more details]{Zuntz2017catalogues}.
We add the redshift calibration shifts calculated by \citep{Hoyle2018redshift}.
The mean redshift of this sample is $\bar z= 0.591$, which is similar to the one obtained with MCCL.
\textsc{Im3shape} and MCCL used different galaxy selection, with roughly 70\% galaxies overlapping between these two catalogues.
The difference in the selection can be attributed to cuts on the PSF size, which is different in the two bands.
Moreover, different bands were used: \mbox{$r$-band} for \textsc{Im3shape} and \mbox{$i$-band} for MCCL, which results in different pixel noise.

We create shear maps with the \textsc{Im3shape} catalogue by pixelising the shapes using equation 7.3 in \citep{Zuntz2017catalogues} and calculate the $C_{\ell}$.
We find that it largely agrees with the one obtained by MCCL, with the average difference of 3.5\% on the $C_{\ell}$, which would correspond to $\approx$1.75\% multiplicative shear bias $m$ if the $n(z)$ were identical.
This level of difference is within our uncertainty on $m$.
We create the covariance matrix for the non-tomographic \textsc{Im3shape} $C_{\ell}$ using the same method as for the MCCL $C_{\ell}$, using the \textsc{Im3shape} $n(z)$ and shape noise.
We marginalise over the redshift and multiplicative shear errors by creating 30 $C_{\ell}$ and $n(z)$ pairs, similarly as in the MCCL method.
For each pair, we modify the $C_{\ell}$ to account for the shear calibration error drawn from the Gaussian prior with $\sigma=0.025$ \citep{Zuntz2017catalogues}, and a shifts drawn from a the prior on $\Delta z$  \citep{Hoyle2018redshift} for each redshift bin, during the creation of the non-tomographic $n(z)$.
With that configuration, we obtain
$S_8 D_{\rm{IA}}=0.857^{+0.063}_{-0.048}$.
With \textsc{Halofit}, we obtain $S_8 D_{\rm{IA}}=0.846^{+0.054}_{-0.045}$.
This is somewhat higher than the tomographic result in \citep{Troxel2017cosmological}, at the level of $\approx 1\sigma$, and lower than the MCCL (non-tomographic) analysis.
This indicates that the specific analysis choice of the DES Y1 data, using non-tomographic $C_{\ell}$ in the range $\ell \in [200, 950]$, gives somewhat higher $S_8 D_{\rm{IA}}$ values than that of the tomographic, real space analysis in \citep{Troxel2017cosmological}.
This level of differences is not unexpected, as selecting different scales for the analysis can cause shifts of the constraints within the statistical uncertainty.

\section{Implementation and runtime}
\label{sec:runtime}
The integrated MCCL pipeline is designed to achieve a fast, integrated analysis.
Our optimized implementation allows to run multiple full-area \band-band simulations in the control loops framework within less than 48 hours.
This process includes: simulating \nsurveys\ full-area DES Y1 simulations from ABC posterior points, measurement of the $n(z)$ curves and shear calibration parameters, calculation of power spectra and covariance matrix, and calculation of the cosmological constraints.
The simulation of a single DES Y1 survey area in the \band-band takes about 2 hour on 400 cores.
Several elements of the pipeline are pre-computed: the ABC posterior on astrophysical and instrument parameters, the trained CNN for PSF estimation and the \textsc{L-Picola} simulations.
From the above, the most time consuming part is the ABC posterior calculation, which takes several days on 1000 cores.
The fast integrated analysis pipeline, which spans the analysis from image pixels to the cosmological constraints, allows for rapid testing of various parameter configurations
and their direct impact on resulting cosmology measurement.

\section{Conclusions}
\label{sec:conclusions}

We have presented the application of the MCCL to a non-tomographic cosmic shear re-analysis of the DES Y1 data.
Using this method, we simultaneously measure the shear E-mode angular power spectrum $C_\ell$ and the redshift distribution $n(z)$ of the galaxy samples used for the shape measurements.

After giving an overview of the method, we discussed the specific implementation for this data set.
In particular, we developed and applied detailed systematic maps for the DES Y1 co-adds.
We also modelled the PSF using deep learning, followed by an interpolation scheme.
We then performed a first control loop to match the simulation to the real data and find a good match following an ABC inference for the simulation parameters.
In the next control loop, we calibrate the shear measurement at the 1-point level and test it using a series of 2-point statistics.
In the third loop, we perform a tolerance analysis for the cosmic shear measurement and derive statistical and systematic uncertainties for both $C_\ell$ and $n(z)$.

Given the resulting set of measurements for these two quantities and after unblinding, we derive cosmological parameter constraints and give an error budget.
Including both statistical and systematic uncertainties, we find
$S_8 D_{\mathrm{IA}}=\sigma_8(\Omega_m/0.3)^{0.5} D_{\mathrm{IA}} = 0.895^{+0.054}_{-0.039}$,
where \defDIA\ is a factor that scales $\sigma_8$ according to the intrinsic alignment amplitude (see Section~\ref{sec:IA}).
We find that the systematic uncertainties contribute to the error budget on the level of $\approx 60\%$ of the statistical error.
Given that our non-tomographic analysis does not lift the degeneracy between $S_8$ and $A_{\rm{IA}}$, a direct comparison with the earlier tomographic analysis \citep{Troxel2017cosmological} is not straightforward.
Considering the degenerate product $S_8 D_{\rm{IA}}$ our results yield a somewhat higher value of this parameter than can be anticipated from this earlier analysis.
To investigate the source of this difference we analyse the DES Y1 \textsc{Im3shape} catalogue using our choices of data vector (non-tomographic $C_{\ell} \in [200, 950]$) and the MCCL likelihood pipeline, using the \textsc{HMcode} power spectrum.
We find a similar non-tomographic $n(z)$ between the MCCL and \textsc{Im3shape}, as well as $C_{\ell}$ in agreement on the 3.5\% level.
The measurement of $S_8D_{\rm{IA}}$ with \textsc{Im3shape} with these settings is close to the MCCL results, and differs on $< 1\sigma$ level.
This indicates that the specific analysis choice we made for the DES Y1 data yields a higher $S_{8}D_{\rm{IA}}$ result than expected from the tomographic analysis of \citep{Troxel2017cosmological}.
The remaining difference can be attributed to different pixel noise between $r$ and $i$-bands, as well as differences in galaxy selection (see Section~\ref{sec:comparison_des}).
Moreover, we find an additional difference in measured $S_{8}D_{\rm{IA}}$ stemming from the use of the \textsc{HMcode} instead of \textsc{Halofit}, on the level of $0.5\sigma$.

The analysis presented here is the first end-to-end application of the MCCL method on a full data set and demonstrates the feasibility and accuracy of the method.
In particular, the fast implementation and integrated nature of the MCCL pipeline offers a very short turnover time thus enabling the exploration, correction and quantification of systematics.
This offers excellent prospects for the application of the MCCL to future weak lensing data sets.
It can also be naturally extended to a tomographic configuration.

\begin{acknowledgments}
We thank Joel Berg\'{e}, Chihway Chang and Lukas Gamper for crucial contributions to the development of \textsc{UFig} and the MCCL method.
We thank Janis Fluri for helpful conversations and help with deep learning aspect of PSF modelling.
We thank Uwe Schmitt and Jarunan Panyasantisuk for informatics support.
We acknowledge support by grant number 200021\_169130 from the Swiss National Science Foundation.
We acknowledge the support of Euler cluster by High Performance Computing Group from ETHZ Scientific IT Services, as well as the support of the Piz Daint cluster by the Swiss National Supercomputing Centre (CSCS).

Funding for the DES Projects has been provided by the U.S. Department of Energy, the U.S. National Science Foundation, the Ministry of Science and Education of Spain,
the Science and Technology Facilities Council of the United Kingdom, the Higher Education Funding Council for England, the National Center for Supercomputing
Applications at the University of Illinois at Urbana-Champaign, the Kavli Institute of Cosmological Physics at the University of Chicago,
the Center for Cosmology and Astro-Particle Physics at the Ohio State University,
the Mitchell Institute for Fundamental Physics and Astronomy at Texas A\&M University, Financiadora de Estudos e Projetos,
Funda{\c c}{\~a}o Carlos Chagas Filho de Amparo {\`a} Pesquisa do Estado do Rio de Janeiro, Conselho Nacional de Desenvolvimento Cient{\'i}fico e Tecnol{\'o}gico and
the Minist{\'e}rio da Ci{\^e}ncia, Tecnologia e Inova{\c c}{\~a}o, the Deutsche Forschungsgemeinschaft and the Collaborating Institutions in the Dark Energy Survey.

The Collaborating Institutions are Argonne National Laboratory, the University of California at Santa Cruz, the University of Cambridge, Centro de Investigaciones Energ{\'e}ticas,
Medioambientales y Tecnol{\'o}gicas-Madrid, the University of Chicago, University College London, the DES-Brazil Consortium, the University of Edinburgh,
the Eidgen{\"o}ssische Technische Hochschule (ETH) Z{\"u}rich,
Fermi National Accelerator Laboratory, the University of Illinois at Urbana-Champaign, the Institut de Ci{\`e}ncies de l'Espai (IEEC/CSIC),
the Institut de F{\'i}sica d'Altes Energies, Lawrence Berkeley National Laboratory, the Ludwig-Maximilians Universit{\"a}t M{\"u}nchen and the associated Excellence Cluster Universe,
the University of Michigan, the National Optical Astronomy Observatory, the University of Nottingham, The Ohio State University, the University of Pennsylvania, the University of Portsmouth,
SLAC National Accelerator Laboratory, Stanford University, the University of Sussex, Texas A\&M University, and the OzDES Membership Consortium.

Based in part on observations at Cerro Tololo Inter-American Observatory, National Optical Astronomy Observatory, which is operated by the Association of
Universities for Research in Astronomy (AURA) under a cooperative agreement with the National Science Foundation.

The DES data management system is supported by the National Science Foundation under Grant Numbers AST-1138766 and AST-1536171.
The DES participants from Spanish institutions are partially supported by MINECO under grants AYA2015-71825, ESP2015-66861, FPA2015-68048, SEV-2016-0588, SEV-2016-0597, and MDM-2015-0509,
some of which include ERDF funds from the European Union. IFAE is partially funded by the CERCA program of the Generalitat de Catalunya.
Research leading to these results has received funding from the European Research
Council under the European Union's Seventh Framework Program (FP7/2007-2013) including ERC grant agreements 240672, 291329, and 306478.
We  acknowledge support from the Brazilian Instituto Nacional de Ci\^encia
e Tecnologia (INCT) e-Universe (CNPq grant 465376/2014-2).

This manuscript has been authored by Fermi Research Alliance, LLC under Contract No. DE-AC02-07CH11359 with the U.S. Department of Energy, Office of Science, Office of High Energy Physics. The United States Government retains and the publisher, by accepting the article for publication, acknowledges that the United States Government retains a non-exclusive, paid-up, irrevocable, world-wide license to publish or reproduce the published form of this manuscript, or allow others to do so, for United States Government purposes.
\end{acknowledgments}

\appendix

\section{Blinding Scheme}
\label{appendix:blinding}
We adopt a blinding scheme to avoid confirmation biases.
For this purpose, we introduce a blinding normalisation and tilt in the EE weak lensing power spectrum $C_{\ell}$.
These two parameters are kept blinded until the results pass predefined tests.

To blind a $C_{\ell}$, we transform it the following way:
\begin{equation}
    C_{\ell}^{\mathrm{blinded}} = C_{\ell} \left( c_0 + \mu(\ell/\ell_0-1) \right)
\end{equation}
where $c_0$ and $\mu$ are blinding factors, and $\ell_0=500$.
The blinding factors were unknown to us, and were drawn from an uniform distribution with ranges $c_0 \in [0.8, 1.2]$ and $\mu \in [-0.3, 0.3]$.
This configuration effectively blinds the $C_{\ell}$ amplitude in $40\%$ range, and adds tilt of up to $30\%$.
This blinding scheme was different than the one used in \citep{Troxel2017cosmological}, where shear values were multiplied by a factor of between 0.9 and 1.1.
The allowed modification of the shear 2-pt function was similar for both strategies.

In Section~\ref{sec:blinding} we listed a set of conditions that need to be satisfied before unblinding.
We consider these conditions to be satisfied, below we summarise the sections in the paper which correspond to each of them.

\emph{(Q1)} The histograms of relevant quantities were described in Section~\ref{sec:abc-fits} and Figure~\ref{fig:histograms_des_abc}.
The histograms obtained from the DES data are contained within the ones from simulations for most of their range, except for tails on several panels.
As these tails contain relatively few objects compared to the bulk of the distribution, we do not expect these differences to impact the result.
We find a small difference in the \texttt{FLUX\_RADIUS} parameter.

\emph{(Q2)} The recovery of the input cosmology for fiducial simulation is satisfactory, as described in Appendix~\ref{appendix:sim_vs_sim}.

\emph{(Q3)} According to the systematics model presented in Section~\ref{sec:systematics_model}, the contribution of remaining systematic maps due to PSF leakage and PSF ellipticity and size modelling errors should not significantly affect the measurement.

\emph{(Q4)} We do not detect significant B-mode (see Section~\ref{sec:2pt_statistics}, which satisfies this condition.

\emph{(Q5)} We test one extension to the model in Section~\ref{sec:loop32}.
We checked if adding scatter to the S\'ersic index of blue galaxies changes the measurement.
We found that neither the shear calibration parameters nor the $n(z)$ change significantly.

The following actions were taken after unblinding.
We found and corrected a mistake in calculation of the GI contribution to the power spectrum.
This change results in an increase in measured $S_8 D_{\rm{IA}}$ by $\approx 0.5 \sigma$.
We also found two minor mistakes in the calculation of covariance matrix concerning the subtraction of the mean of the convergence maps and rotation of the shear for some of the ten patches.
As the covariance matrix is dominated by shape noise, we found that these mistakes did not noticeably affect the final result.

\section{Power spectrum calculation}
\label{appendix:power_spectrum_calculation}

\begin{figure*}[t!]
\includegraphics[width=1.0\textwidth]{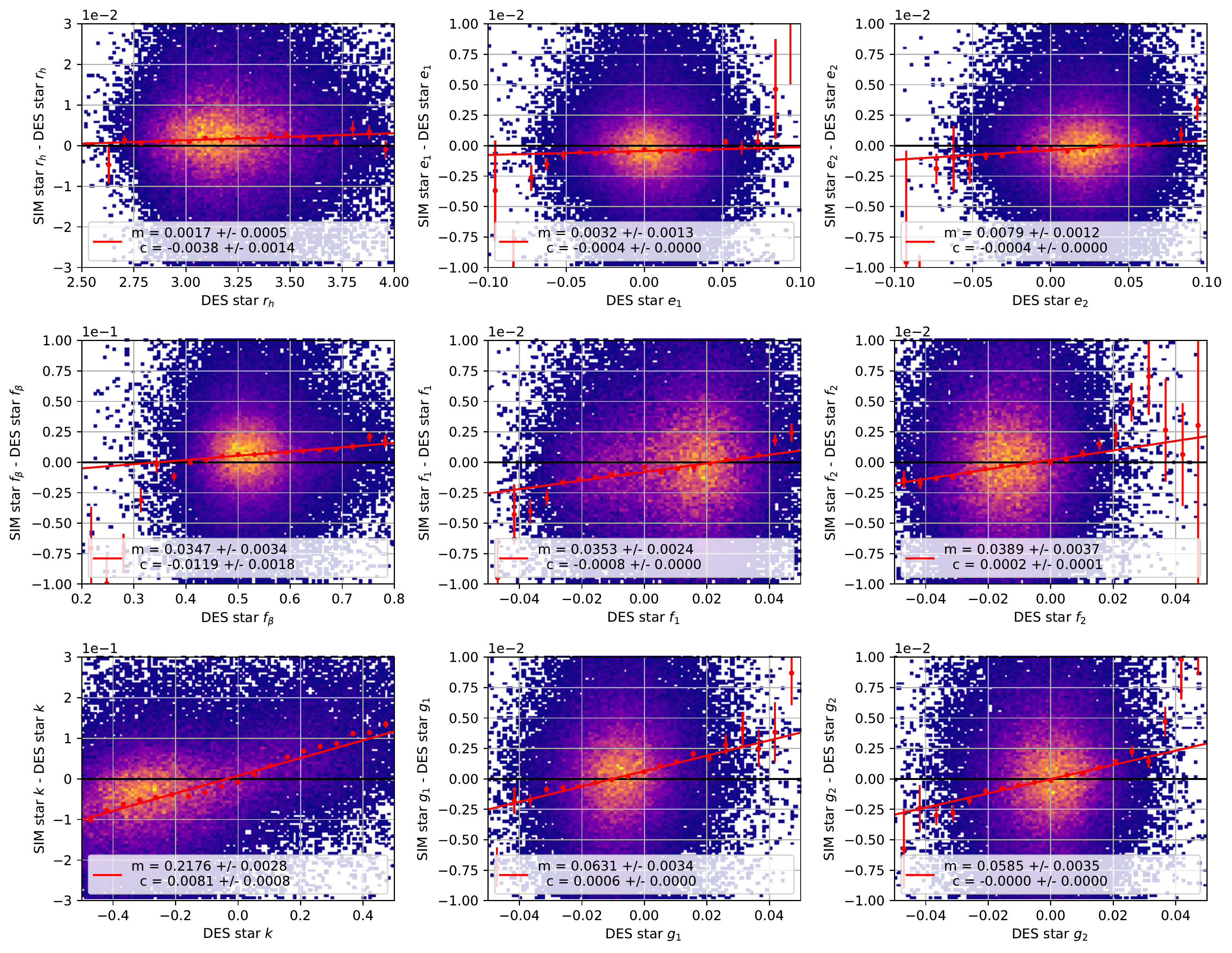}
\caption{Accuracy of the PSF measurement and interpolation for PSF model parameters: size $r_p$, ellipticities $e_i$ flexions $f_i$, $g_i$, flux ratio of profile components $f_{\beta}$, and kurtosis $k$.
Each panel shows the residual between the interpolated value and the true value of the parameter as a function of the true value.
Red points correspond to the mean of the residual in bins, and the error-bars represent the standard deviation.
Red lines show the line fit to these points, with parameters stated in the legend, where $m, c$ are the slope and intercept, respectively.
}
\label{fig:psf_1pt_in_vs_out}
\end{figure*}

We create pixelised shear maps from the source galaxy sample by averaging the shapes of all galaxies in each pixel, with no weighting of shapes applied.
We use the HEALPix scheme to create the maps.
We compute all spherical harmonic power spectra using the publicly available code \texttt{PolSpice}\footnote{http://www2.iap.fr/users/hivon/software/PolSpice/} \cite{Szapudi:2001, Szapudi:2001ab, Chon:2004}. In order to estimate the values of the maximal angular scale used by $\tt{PolSpice}$, $\theta_{\mathrm{max}}$, and the apodization parameter $\theta_{\mathrm{FWHM}}$, we follow the method outlined in ~\cite{Nicola:2016, Nicola:2017}.
From this stage onwards, we follow the blinding scheme described in Appendix~\ref{appendix:blinding}.
Calculated $C_{\ell}$ were averaged into 15 linearly-spaced bins in the range between $\ell_{\mathrm{min}}=200$ and $\ell_{\mathrm{max}}=950$.

For the ellipticity maps, the mean of the map, corresponding to $\ell=0$, is subtracted before passing it to $\textsc{PolSpice}$.
We process scalar quantities, such as PSF size, in the form of fractional difference maps, $(x-\bar x)/(\bar x)$, where $\bar x$ is the mean of the map.

We subtract the contribution of the noise from auto power spectra.
The noise contribution is calculated by taking a mean of 100 $C_{\ell}$ calculated from maps with removed signal.
For spin-2 ellipticities, the signal was removed by randomly rotating the shapes.
For scalar maps, the positions of the stars/galaxies were randomly permuted.

To account for variable number of galaxies in each pixel, we compute the power spectrum using an additional weight map.
The weight map corresponds to inverse variance weights and is simply a map of source galaxy number counts.

\section{PSF estimation and modelling}
\label{appendix:psf}

As described in Section~\ref{sec:loop1}, we rely on the approach presented by \citep{Herbel2018Fast} to estimate and model the PSF.
However, we modify the parametrisation of the  PSF model, parameter estimation details, and neural network training strategy.
We detail and motivate these changes in this section.
We also present the details of the interpolation method on co-added images.

\subsection{PSF model and parameter estimation}

The updated model contains a modified base profile.
We use a mixture of two Moffat profiles with $\beta_1=2$ and $\beta_2=5$.
Parameter $f_{\beta} \in [0,1]$ controls the fraction of the photons sampled from the first profile.
Note that in \citep{Herbel2018Fast} this parameter was fixed and labelled $\gamma$.
We also modify the perturbation of the photon position associated with the kurtosis operation.
The position of photons $\theta_i$ is perturbed by $\delta \theta_i$ in the following way:
\begin{align}
\delta \theta_i &= k \cdot \norm{\theta}_2^{2} \cdot \beta^{-1} \cdot \theta_i \cdot k_s, \\
k_s &=\exp \left[ -0.5 \norm{\theta}_2^2/(0.2\beta^{2}) \right].
\end{align}
where $k$ is the kurtosis parameter and $k_s$ is a suppression factor dependent on the profile.
This factor prevents extremely large displacements for photons in the wings of the profile.
In total, the model has 11 parameters: centroid positions $x$, $y$, full width half maximum (FWHM) $r_{h}$, ellipticities $e_1$, $e_2$, flexions $f_1$, $f_2$, $g_1$, $g_2$, flux ratio between two Moffat components $f_{\beta}$ and kurtosis $k$.
The centroid positions are not used later in any way.
We found that it is difficult to measure the kurtosis parameter, as it does not modify the profile in a significant way within the prior range we considered.

We used image cutouts of size $19 \times 19$ pixels.
To generate the training data, we draw magnitudes between 17 and 22, gain values between 3 and 5.5$\, e^- / \text{ADU}$, the number of exposures between 1 and 9 and we use a magnitude zeropoint of 30, which is the value of the DES Y1 data.
During the training, we add Gaussian background noise with standard deviations sampled from the interval $0 \, \text{ADU}$ to $10 \, \text{ADU}$ on the fly.

We train the CNN using a different objective function than \citep{Herbel2018Fast}, the likelihood loss, similarly to \citep{Fluri2018deeplearning}.
We use single parameter variances instead of a full covariance matrix.
Another modification to the cost function is the use of averaging of multiple noise realisations before taking the square of the residuals, which greatly reduced the bias of the recovered parameters.
The cost $L$ is:

\begin{equation}
L = \sum_{ i \in N_b} \sum_{p \in N_p} \left[ \left\langle \frac{\hat \theta_{i,n}^{p} - \theta_{i,n}^{p}}{\sigma_{p}} \right\rangle_{N_n}^{2}
+   \log{ \left\langle  \sigma_p \right\rangle }^{2}_{N_n} \right],
\end{equation}
where $N_b=2048$ is the number of images in a batch, $N_p=11$ is the number of parameters in the model, $N_n=64$ is the number of noise realisations per parameter set,
$\hat \theta_{i,n}^{p}$ is the network's prediction, $\theta_{i,n}^{p}$ is the true parameter and $\sigma_{p}$ is the uncertainty of parameter $p$, the logarithm of which is also predicted by the network.
Angular brackets denote averaging over the number of noise realisations $N_n$.
To train using this loss function, we generated a training set consisting of $10^{9}$ star images samples using Latin Hypercube Sampling in ranges of magnitude, gain and number of exposures  mentioned above. The remaining parameters had the following span:
$r_{h}        \in [2.5, 5]$,
$e_1, e_2, f_1, f_2, g_1, g_2 \in [-0.1, 0.1]$,
$k    \in [-0.5, 0.5]$,
$x, y               \in [9, 10]$,
$f_{\beta}  \in [0., 1.]$.

In the training process we progressively increase the noise levels and decreased the learning rate.
We trained the network with no added Gaussian noise (the training images already contained Poisson noise) on 100'000 iterations, then we gradually increase the noise level until it reaches $\sigma_n=10$ at 200'000 iteration.
The learning rate is initiated at $l_r=0.001$, decreased at steps [$5\cdot10^{4}$, $10^{5}$, $2\cdot10^{5}$] to $l_r=[10^{-4}, 5\cdot10^{-5}, 2\cdot10^{-5}]$, respectively.
We stored the network that gave us the best loss value, which happened at iteration 562,186.

\subsection{PSF interpolation}

We interpolate the PSF across the co-added images.
We model the PSF surface for each of the parameters, except the $x,y$ positions, using a Chebyshev polynomial basis of maximum order 4.
The basis also includes information about the tiling pattern coming from the co-adds.
We create a pseudo-Vandermonde matrix $\Phi_{ijk}$ of degree 4 for sample points $x$ and $y$ in the co-add pixel coordinate system
\begin{equation}
    \Phi_{ijk}(x, y) = T_{i}(x)  T_{j}(y) W_k(x, y),
    \label{eqn:psf_interpolation_basis}
\end{equation}
where $T_{i}$ is the Chebyshev polynomial of order $i$ and $W_k$ is a weight corresponding to exposure $k$.
If an object at position $(x,y)$ was recorded by exposure $k$, then the weight is non-zero, and has a value $W_k(x, y) = 1/N^{e}_{x,y}$, where $N^{e}_{x,y}$ is the total number of exposures contributing to the image at position $(x,y)$.
This model assumes equal-weighted contributions to the PSF parameters from each exposure, which may not be true in general; we do not attempt to make a more detailed model of this process, leaving that to future work.
The basis in Equation \ref{eqn:psf_interpolation_basis} has total of $4^2 N_{\rm{exp}}$ parameters for each co-add, where $N_{\rm{exp}}$ is the maximum number of exposures for that co-add.
We fit a surface for each parameter iteratively, using a robust technique based on $\sigma$-clipping.
After the initial fit, the $3\sigma$ outliers in $r_p, e_1, e_2$ are removed, and the fit is performed again until convergence, and at maximum 10 times.
The final model is stored.
The iterative fitting procedure makes the model more robust to stars with unusual parameters, extreme measurement errors, stars blended with galaxies and false positives in the \textit{Gaia} catalogue.
Before fitting, we exclude 15\% of stars randomly to create a validation star sample, used later for testing the PSF interpolation.

Figure \ref{fig:psf_1pt_in_vs_out} shows the accuracy of the PSF modelling from our fiducial simulation.
We plot the difference between PSF parameters estimated from $\sim 60,000$ validation stars in DES and simulations at the exact same positions.
The errors in predictions are shown as a function of the parameter estimated for the DES star.
The red points show the means of residuals in bins spanning the range shown, and the error-bars correspond to error on the mean.
The red line shows a linear fit to these points.
We notice that the PSF recovery is generally good; small biases exist for size, ellipticities and flexions, while the recovery is worse for kurtosis and profile flux ratio.
Similar level errors is observed when the full sample of stars is used, including the ones used for PSF estimation, as well as in simulations, when the estimated PSF model parameters are compared with the true input.
As the overall error in E-mode reconstruction in simulations is satisfactory with respect to the requirements (see Appendix~\ref{appendix:sim_vs_sim}), we conclude that these differences do not have a significant impact on the shear calibration.

\section{ABC analysis}
\label{appendix:abc-analysis}

\begin{figure*}[t!]
\includegraphics[width=1\textwidth]{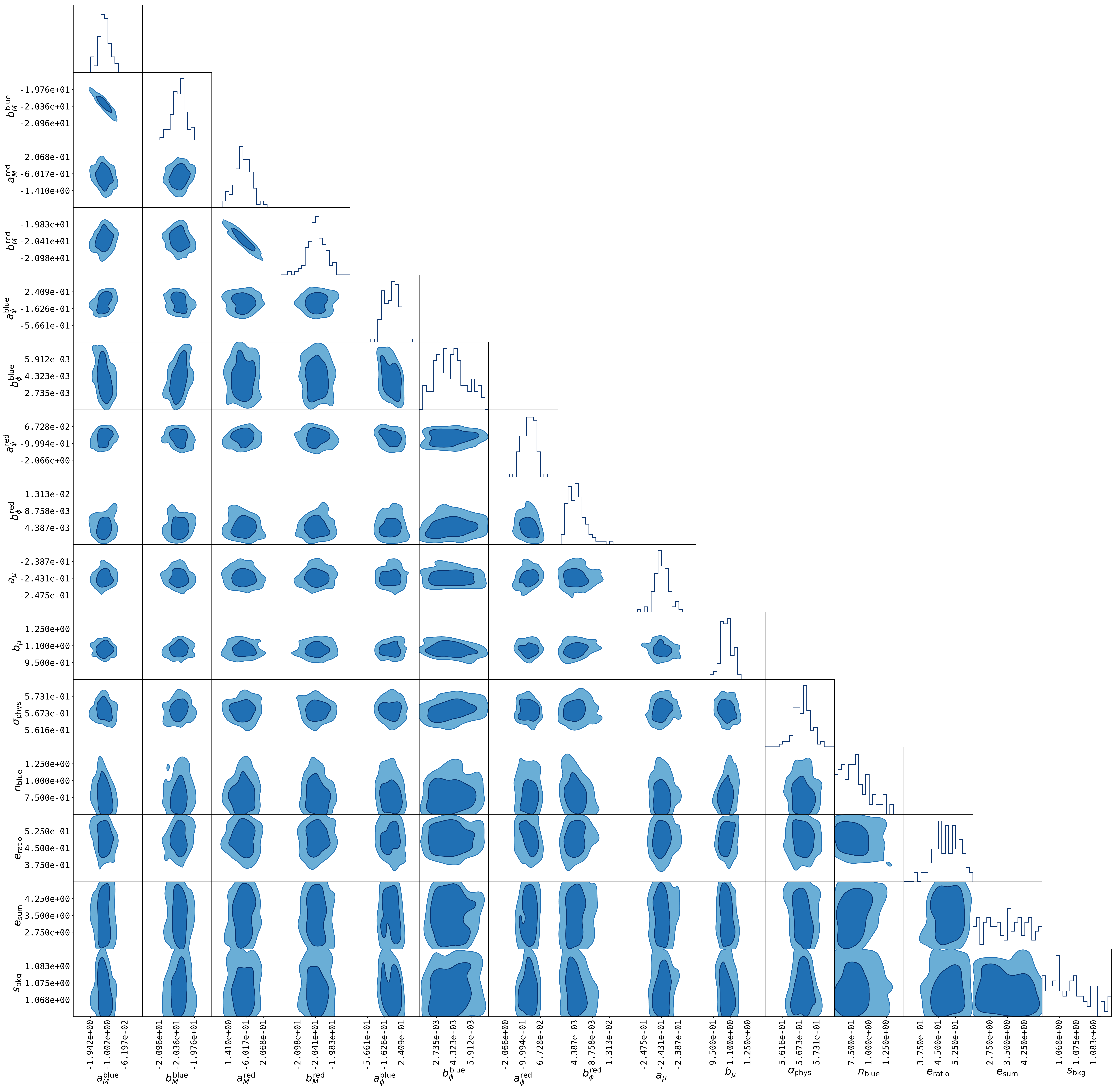}
\caption{Posterior distribution on the parameters of the forward model.
This distribution was found using ABC.
Parameters of the template coefficients are not shown.
The shaded regions show 68\% and 95\% confidence intervals.
}
\label{fig:abc_posterior_samples}
\end{figure*}

\subsection{Variable parameters \& priors}
\label{sec:variable-parameters-priors}

In this section, we present more details on the parameters we vary during the ABC analysis and specify the priors used the sample the 35-dimensional parameter space. As described in section \ref{sec:abc-fits}, the parameters we vary can be divided into six groups, which are listed below. The combined prior is a product of the individual priors. Table \ref{tab:abc-parameters} summarizes the information presented in this section.

\begin{enumerate}[wide, labelwidth=!, labelindent=0pt]

\item
\textit{Galaxy luminosity function parameters.} These parameters are referred to as $a_M$, $b_M$, $a_\phi$, $b_\phi$ in \cite{Herbel2017redshift} for the blue and the red population, respectively. They control the redshift evolution of the luminosity functions from which we sample the galaxy population. We use a modified version of the prior specified in \cite{Herbel2017redshift}, which was sampled with \textsc{CosmoHammer} using data given in \cite{Beare2015}. Since this prior is well described by a multivariate normal distribution, we use the mean and covariance matrix computed from the \textsc{CosmoHammer} samples used by \cite{Herbel2017redshift} to generate the new prior data for this work, which simplifies the sampling. We retained the enlarged prior volumes used in \cite{Herbel2017redshift} for $b_\phi$. However, we reduced the enlargement for blue galaxies by multiplying each sample with a random number between $0.5$ and $2.5$ (instead of $0.5$ and $4$, as in \cite{Herbel2017redshift}). Furthermore, we decided to limit the range of $b_\phi$ to the interval $[0, 0.0075]$ for blue galaxies and to $[0, 0.0175]$ for red galaxies. If a sample does not fall within these boundaries, we redraw. These modification were inspired by earlier ABC analyses, they are effectively a by-hand application of the ABC algorithm proposed by \cite{Kacprzak2017}.

\item
\textit{Galaxy SED parameters.} The parameters $\alpha_{i,0}$ (at redshift $z=0$) and $\alpha_{i,1}$ (at redshift $z=1$) control the Dirichlet distributions from which we sample the template coefficients used to assign restframe SEDs to galaxies, see \cite{Herbel2017redshift}. We retain the prior used by \cite{Herbel2017redshift} for this work.

\item
\textit{Galaxy size parameters.} The parameters $a_\mu$, $b_\mu$, $\sigma_\text{phys}$ from \cite{Herbel2017redshift} set our model for the distribution of intrinsic galaxy sizes. As for the galaxy luminosity function parameters, we sample this part of parameter space using a multivariate normal distribution with the mean and covariance matrix computed from the corresponding \texttt{CosmoHammer} samples used by \cite{Herbel2017redshift}. Since $\sigma_\text{phys}$ denotes a standard deviation, this parameter cannot be negative, therefore, we redraw if this happens. Furthermore, as was done in \cite{Herbel2017redshift}, we relax the prior on $b_\mu$. Here, we use a uniform prior ranging from 0.8 to 1.4.

\item
\textit{S\'ersic index for blue galaxies.} We sample $n_\text{blue}$ from a uniform prior extending from 0.5 to 1.5.

\item
\textit{Intrinsic ellipticity distribution for galaxies.} The uniform prior for $e_{\rm{ratio}}$ ranges from 0.3 to 0.6 and the uniform prior for $e_{\rm{sum}}$ from 2 to 5.

\item
\textit{Background noise scale.} This parameter scales the background level noise globally for all simulated images. We place a uniform prior extending from 1.06 to 1.09 on this parameter.

\end{enumerate}

\begin{table*}
\medmuskip=0mu
\setlength{\tabcolsep}{5pt}
\footnotesize
\centering
    \begin{tabular}{ p{0.02\linewidth} | p{0.07\linewidth} | p{0.55\linewidth} | p{0.3\linewidth} }
        \hline
        & Parameter & Meaning & Prior \\
        \hline
        \multirow{8}{*}[-14mm]{\begin{sideways}Luminosity functions\end{sideways}} & $a_M^\mathrm{blue}$ & Slope of the redshift of evolution of the parameter $M_*$ in the Schechter luminosity function for blue galaxies & Multivariate normal, $\mu=-9.43 \cdot 10^{-1}$, $\sigma^2=2.09 \cdot 10^{-1}$ \\
        & $b_M^\mathrm{blue}$ & Intercept of the redshift of evolution of the parameter $M_*$ in the Schechter luminosity function for blue galaxies & Multivariate normal, $\mu=-2.04 \cdot 10^{1}$, $\sigma^2=8.33 \cdot 10^{-2}$ \\
        & $a_M^\mathrm{red}$ & Slope of the redshift of evolution of the parameter $M_*$ in the Schechter luminosity function for red galaxies & Multivariate normal, $\mu=-7.07 \cdot 10^{-1}$, $\sigma^2=1.35 \cdot 10^{-1}$ \\
        & $b_M^\mathrm{red}$ & Intercept of the redshift of evolution of the parameter $M_*$ in the Schechter luminosity function for red galaxies & Multivariate normal, $\mu=-2.04 \cdot 10^{1}$, $\sigma^2=7.46 \cdot 10^{-2}$ \\
        & $a_\phi^\mathrm{blue}$ & Decay constant of the redshift of evolution of the parameter $\phi_*$ in the Schechter luminosity function for blue galaxies & Multivariate normal, $\mu=-1.17 \cdot 10^{-1}$, $\sigma^2=3.49 \cdot 10^{-2}$ \\
        & $b_\phi^\mathrm{blue}$ & Amplitude of the redshift of evolution of the parameter $\phi_*$ in the Schechter luminosity function for blue galaxies & Multivariate normal, $\mu=3.76 \cdot 10^{-3}$, $\sigma^2=2.29 \cdot 10^{-7}$, multiplied by random variate from $[0.5, 2.5]$, bounded to $[0, 7.5 \cdot 10^{-3}]$ \\
        & $a_\phi^\mathrm{red}$ & Decay constant of the redshift of evolution of the parameter $\phi_*$ in the Schechter luminosity function for red galaxies & Multivariate normal, $\mu=-8.96 \cdot 10^{-1}$, $\sigma^2=2.47 \cdot 10^{-1}$ \\
        & $b_\phi^\mathrm{red}$ & Amplitude of the redshift of evolution of the parameter $\phi_*$ in the Schechter luminosity function for red galaxies & Multivariate normal, $\mu=3.91 \cdot 10^{-3}$, $\sigma^2=1.14 \cdot 10^{-6}$, multiplied by by random variate from $[0.5, 4]$, bounded to $[0, 1.75 \cdot 10^{-2}]$ \\
        \hline
        \multirow{3}{*}[-6mm]{\begin{sideways}Galaxy sizes\end{sideways}} & $a_\mu$ & Slope of the evolution of the average intrinsic physical size of galaxies with absolute magnitude & Multivariate normal, $\mu=-0.24 \cdot 10^{0}$, $\sigma^2=3.31 \cdot 10^{-6}$ \\
        & $b_\mu$ & Intercept of the evolution of the average intrinsic physical size of galaxies with absolute magnitude & Uniform in $[0.8, 1.4]$ \\
        & $\sigma_\mathrm{phys}$ & Standard deviation of the normal distribution we use to sample intrinsic physical galaxy sizes & Normal multivariate, $\mu=0.57 \cdot 10^{0}$, $\sigma^2=6.36 \cdot 10^{-6}$, bounded to positive values \\
        \hline
        \parbox[c]{0mm}{\rotatebox[origin=c]{90}{\parbox[c]{11mm}{\tiny Galaxy \\ profiles}}} & $n_\mathrm{blue}$ & S\'ersic index of blue galaxies & Uniform in $[0.5, 1.5]$ \\
        \hline
        \multirow{2}{*}[-0.5mm]{\tiny \begin{sideways}\shortstack[c]{Galaxy \\ ellipticities}\end{sideways}} & $e_\mathrm{ratio}$ & \multirow{2}{*}{\shortstack[l]{Parameters controlling the beta distribution from which we sample intrinsic \\ galaxy ellipticities}}  & Uniform in $[0.3, 0.6]$ \\
        & $e_\mathrm{sum}$ & & Uniform in $[2, 5]$ \\
        \hline
        \parbox[c]{0mm}{\rotatebox[origin=c]{90}{\parbox[c]{14.5mm}{\tiny Background \\ noise}}} & $s_\mathrm{bkg}$ & Scale factor for the background noise level in the simulations & Uniform in $[1.06, 1.09]$ \\
        \hline
        \multirow{4}{*}[-6mm]{\begin{sideways}Template coefficients\end{sideways}} & $\alpha^\mathrm{blue}_{i,0}$ & Concentration parameters of the Dirichlet distribution at redshift $z=0$ from which the template coefficients for blue galaxies are sampled, $i=1,...,5$ & \multirow{4}{*}[-7mm]{\shortstack[l]{Dirichlet distributions with equal concen-\\tration parameters; the sums of the con-\\centration parameters are uniformly\\ distributed in $[5, 15]$}} \\
        & $\alpha^\mathrm{blue}_{i,1}$ & Concentration parameters of the Dirichlet distribution at redshift $z=1$ from which the template coefficients for blue galaxies are sampled, $i=1,...,5$ & \\
        & $\alpha^\mathrm{red}_{i,0}$ & Concentration parameters of the Dirichlet distribution at redshift $z=0$ from which the template coefficients for red galaxies are sampled, $i=1,...,5$ & \\
        & $\alpha^\mathrm{red}_{i,1}$ & Concentration parameters of the Dirichlet distribution at redshift $z=1$ from which the template coefficients for red galaxies are sampled, $i=1,...,5$ & \\
        \hline
   \end{tabular}
   \caption{Variable parameters and priors for the ABC analysis. See section \ref{subsec:galaxy-population-model}, appendix \ref{sec:variable-parameters-priors} as well as \cite{Herbel2017redshift} for more information.
   The values of $\sigma^2$ for the multivariate normal distributions correspond to the diagonal element of the full covariance matrix.
   }
   \label{tab:abc-parameters}
\end{table*}

\subsection{Distance metrics}
\label{sec:distance-metrics}

We use the five distance metrics listed below to infer our ABC posterior. As described in section \ref{sec:abc-fits}, we evaluate all samples on 20 DES tiles and average the individual values of the distance metrics to reduce cosmic variance.

\begin{enumerate}[wide, labelwidth=!, labelindent=0pt]
\item \textit{Magnitude histogram distance.}
This distance metric ensures that our posterior simulations match the DES data in terms of galaxy number counts and $r$-band magnitude distributions.
It is computed by subtracting the binned $r$-band magnitude distribution of the DES lensing galaxies from the magnitude distribution of the simulated lensing galaxies and summing up the absolute differences, see \cite{Herbel2017redshift}.
This distance metric is averaged by stacking the histograms from the individual tiles and averaging the bin entries.
The distance metric is then evaluated on the averaged histograms.

\item \textit{MMD distance for magnitudes, sizes and ellipticity.}
We compute a 11-dimensional MMD distance (Maximum Mean Discrepancy, see \cite{Gretton2012, Herbel2017redshift}) using the measured magnitudes and sizes in the five DES filter bands as well as the measured ellipticity in the $r$-band.
This distance metric selects samples which match the DES data in terms of colors, magnitude-size-distributions and \band-band ellipticities.
To evaluate the MMD distance, we use the lensing galaxies.
The values of the distance metrics computed from the individual DES tiles are directly averaged to obtain a mean value for one sample.

\item \textit{VVDS-Wide MMD distance.}
As was done in \cite{Herbel2017redshift}, we use the VVDS-Wide spectroscopic sample to compute a two-dimensional MMD distance using measured $i$-band magnitudes as well as redshifts.
The VVDS-Wide sample is purely magnitude-limited in the $i$-band, such that we can easily emulate that sample from our simulations and use it to tighten our constraints on the redshift distribution.
To average this distance metric over multiple DES tiles, we directly average the distance values obtained from the individual co-added images.

\item \textit{Galaxy profile distance.}
To constrain the S\'ersic index for blue galaxies, we use a distance metric based on the light distributions of our lensing galaxies in the $r$-band.
We cut out $21 \times 21$ pixel stamps of these objects and compute the pixel-wise mean, which yields the average lensing galaxy profile for one tile.
We normalize the profile to have a maximum pixel flux of 1 and then compute the pixel-wise sum of the squared difference of the profile from the simulated data and the one obtained from DES data.
To average this distance metric over multiple tiles, we first average the mean profiles extracted from the individual tiles and then evaluate the distance metric.

\item \textit{Background level distance.}
To constrain the parameter scaling the background level of the simulated images, we use histograms of pixel values between $-10$ and $10\,$ADU, which characterize the background noise.
We compute these histograms from the DES data and from the simulations.
We then subtract the histogram computed from DES from the histogram obtained from a simulated image and sum up the absolute differences in the bin entries to obtain a distance metric.
To average this distance metric over multiple tiles, we stack the corresponding histograms and average the bin entries.
The distance metric is then evaluated on the averaged histograms.

\end{enumerate}

Concerning the thresholding prescription, we adapt the method given by \cite{Kacprzak2017}. We scale all distance metrics to cover similar numerical ranges by dividing by the corresponding 10th percentiles. Furthermore, we downweight distances number 1 and 5 by a factor of 0.2, which results in tighter constraints on $n(z)$ and the size and ellipticity distributions. We then combine all five distance metrics into a single value by taking the maximum. This combined distance metric is subsequently used to select a number of best samples.

\begin{figure*}[t!]
\includegraphics[width=0.7\linewidth]{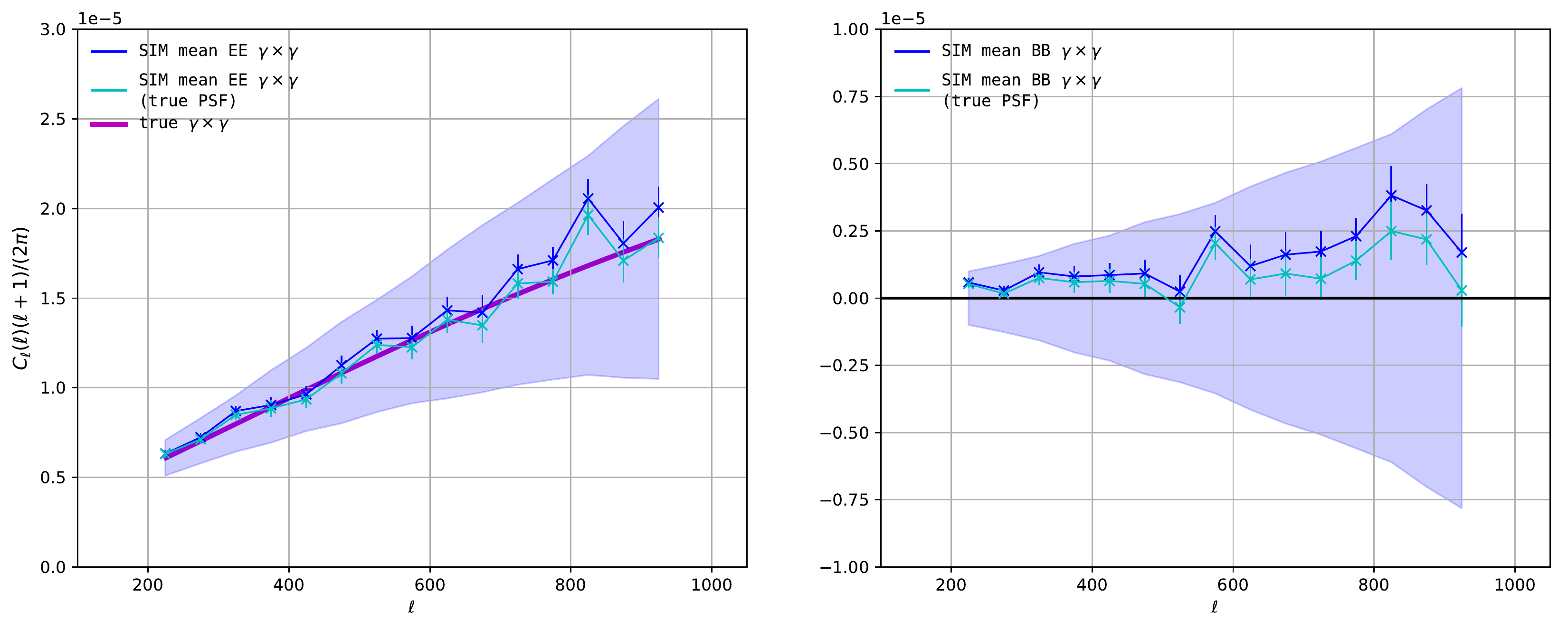}
\caption{E-mode (left) and B-mode (right) average power spectra of \nsurveys\ UFig simulations from the fiducial survey, with different random seeds.
The true $C_{\ell}$ is shown with the magenta line.
The blue (cyan) line shows the power spectrum calculated using estimated (true) PSF parameters.
Error-bars on the lines show the error on the mean from \nsurveys\ power spectra.
Light blue bands correspond to $1\sigma$ errors as calculated from the covariance matrix (see Section~\ref{sec:joint_cal}).
}
\label{fig:2pt_xcorrs_sim_vs_sim}
\end{figure*}

\section{Galaxy sample selection}
\label{appendix:galaxy_sample_selection}

The minimum and maximum galaxy-to-PSF size ratios were set to $r_{g}/r^{p}_{h}>0.75$ and $r_{g}/r^{p}_{h}<100$, respectively,
where
\begin{equation}
r_{g} = \sqrt{ (M_{00}+M_{11}) \cdot 2 \cdot \log(2) }
\label{eqn:galaxy_size}
\end{equation}
is the measured galaxy size  \citep{Bruderer2017shear}, and the elements of the moment matrix $M$ are the \textsc{SExtractor} windowed moments $M_{00}$=\texttt{X2WIN\_IMAGE} and $M_{11}$=\texttt{Y2WIN\_IMAGE}.

The minimum signal-to-noise ratio to $S/N>10$, where $S/N=\texttt{FLUX\_AUTO/FLUXERR\_AUTO}$.
We required the objects in the Galaxy and PSF sample to have \textsc{SExtractor} flags \texttt{FLAGS=0} or \texttt{FLAGS=16}.
This ensured the removal of of blended objects.
We found that excluding objects with \texttt{FLAGS=16} set to 1 was causing large selection biases on the shape.
This is due to \texttt{FLAGS=16} parameter being affected by the row-by-row scanning strategy employed by \textsc{SExtractor}.
We additionally removed all galaxies from the source galaxy sample that had sizes outside the range $r_g \in [2,20]$ and ellipticities of the \textsc{SExtractor} windowed moments $|e_{M}|>1$.
This ellipticity was simply calculated as $e_{M,1}=(M_{00}-M_{11})/(M_{00}+M_{11})$ and $e_{M,2}=(2M_{01})/(M_{00}+M_{11})$.
We also removed galaxies for which the PSF prediction was unreliable, with the criterion: $r_{h} \in [2,12]$ and $|e_{\mathrm{PSF}}|<0.5$.

We removed all objects which lie on the border between chip images.
We required that there are no breaks in the exposure maps (see Section~\ref{sec:systematics_maps}) inside the galaxy's postage stamp of size 20 pixels.
We expect the images lying on the boarder to have unreliable PSF models.

We remove all objects in areas covered by less than 3 exposures.
We found that this cut greatly improves the B-mode statistics.
This is due to the fact that a lot of these objects lie in areas between chips.
These areas can have an unreliable PSF model.
This cut removed around 25\% galaxies.

The analysis used 3373 tiles associated with the \texttt{Y1A1} tag.
However, we removed 18 tiles for which the \textsc{SExtractor} run or PSF estimation was consistently failing.
These tiles were: DES0001-5705, DES0319-6456, DES0030-4331, DES2019-5957, DES2335-5705, DES2008-5248, DES0346-6456, DES0347-4123, DES0339-6039, DES2225-5957, DES0622-5248, DES0434-3957, DES2248-4706, DES2240-4623, DES2244-4706, DES0044-4123, DES0620-5331, DES2312-5123.

The \textsc{SExtractor} parameters use were:
\footnotesize
\verb+DETECT_TYPE=CCD, DETECT_MINAREA=5, DETECT_THRESH=1.7,+ \\
\verb+ANALYSIS_THRESH=1.7, FILTER_NAME=gauss_2.0_5x5.conv,+ \\
\verb+DEBLEND_NTHRESH=32, CLEAN=Y, DEBLEND_MINCONT=0.000005,+ \\
\verb+CLEAN_PARAM=1.0, MASK_TYPE=CORRECT, PHOT_APERTURES=5,+ \\
\verb+FILTER=Y, PHOT_AUTOPARAMS=2.5, 3.5, SATUR_LEVEL=300000,+ \\
\verb+PIXEL_SCALE=0.263, PHOT_FLUXFRAC=0.5, BACK_SIZE=64,+ \\
\verb+BACK_FILTERSIZE=3, BACKPHOTO_TYPE=LOCAL,+ \\
\verb+MEMORY_OBJSTACK=20000, MEMORY_PIXSTACK=50 \\00000,+ \\
\verb+MEMORY_BUFSIZE=2000., BACKPHOTO_THICK=24.+
\normalsize

\section{Simulation internal test}
\label{appendix:sim_vs_sim}

\begin{figure*}
\includegraphics[width=0.7\linewidth]{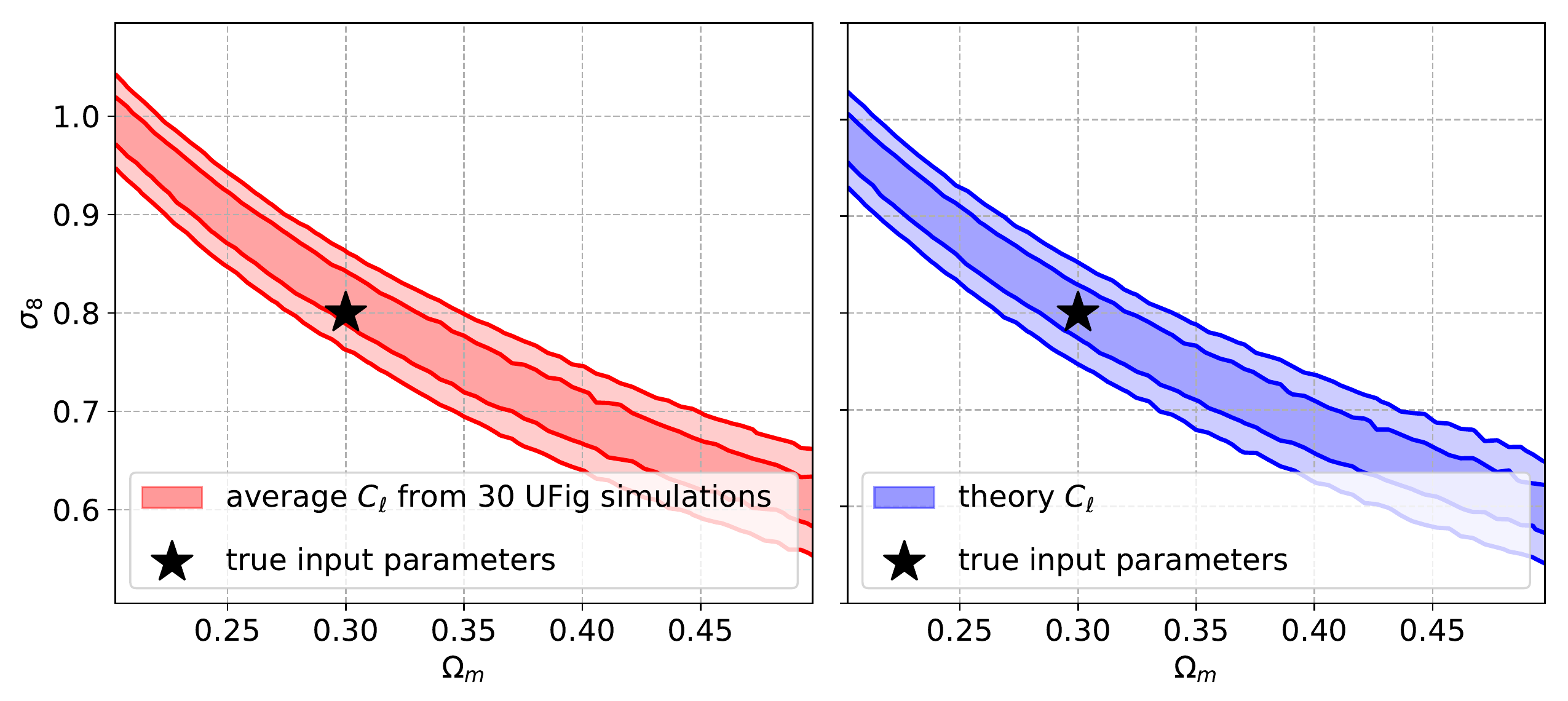}
\caption{
Constraints from average $C_{\ell}$ from \nsurveys\ UFig simulations of full Y1 area with different random seeds (left panel).
The input cosmology, true shear map, and systematic maps are kept the same.
The estimated PSF parameters are used in the $C_{\ell}$ calculation.
The true input cosmology is marked with a black star.
The right panel shows the constraints obtained when the true $C_{\ell}$ is used as input.
The covariance matrix used is the same as in the main analysis.
These constraints contain only the statistical uncertainty, no maringalisation over systematics from $n(z)$ or shear calibration is performed.
}
\label{fig:constratints_sim_vs_sim}
\end{figure*}

We perform an end-to-end test of the analysis pipeline to determine if we can recover the input cosmological parameters.
We simulate a Gaussian shear field using \texttt{Synfast}, a part of \texttt{HEALPy} package, with the input power spectrum corresponding to the parameters $h=0.7$, $\Omega_b=0.05$, $\Omega_m=0.3$, $\sigma_8=0.8$, and an $n(z)$ of the fiducial survey.
We run another \nsurveys\ simulations with the fiducial survey parameters (see Section \ref{sec:abc-fits}), changing only the random seeds in UFig.
The input shear map and systematic maps were the same, while pixel noise, galaxy positions and parameters were randomly drawn from \nsurveys\ different seeds.
We analyse these \nsurveys\ surveys separately in the exact same way as the DES data and apply the calibration parameters from the fiducial survey.
Finally, we average the power spectra after noise correction.
This allows us to reduce the statistical uncertainty of the power spectrum measurement.
Left and right panels on Figure~\ref{fig:2pt_xcorrs_sim_vs_sim} show the average EE and BB $C_{\ell}$.
The true $C_{\ell}$ is shown with magenta line.
The blue line shows the $C_{\ell}$ obtained with estimated PSF parameters used as an input for shear calibration.
The cyan line shows the mean $C_{\ell}$ calculated using the  true PSF parameters.
This is the best-case scenario, when the PSF information is perfectly known.
Error-bars on the lines correspond to the errors on the mean $C_{\ell}$ and they exclude the cosmic variance.
The light blue band corresponds to the $1\sigma$ errors for a single survey, and are taken from the covariance matrix diagonal, similarly to Figure~\ref{fig:shear_bmode}.

The recovery of the power spectrum is generally good for the case when true PSF parameters were used in Equation~\ref{eqn:shear_calibration}.
We notice, however, a slight error on the recovered mean $C_{\ell}$ when the estimated PSF parameters are used.
To examine the impact of that error, we calculated cosmological constraints using the mean $C_{\ell}$ of multiple realisations of the UFig full simulations.
We averaged the \nsurveys\ $C_{\ell}$ and passed it as an input to the likelihood analysis, the same way as for the main result presented in Section~\ref{sec:cosmo_constraints}, including the same covariance matrix.
No baryons or intrinsic alignments were used in this test.
The left panel of Figure~\ref{fig:constratints_sim_vs_sim} presents the $\sigma_8-\Omega_m$ constraints for the average $C_{\ell}$.
The input cosmology is marked with a star.
These constrains do not include marginalisation over the systematic uncertainty from $n(z)$ and shear calibration.
The constraint lies within $0.5\sigma$ away from the input.
The difference would be even less significant if the systematic uncertainty was also marginalised.
The right panel shows the constraints if the true $C_{\ell}$ is used.
We also do not see any significant deviation of the constraint from the truth.
Both experiments indicate that the errors arising from the imprecisions in PSF modelling and interpolation, as well as finite number of simulations used to create the covariance matrix, are not affecting the constraints on a significant level.
We conclude that the analysis pipeline recovers the input cosmological parameters well from internally created simulations.

\section{Shear 1-pt statistics}
\label{appendix:shear_1pt}

\begin{figure*}[t!]
\includegraphics[width=1\linewidth]{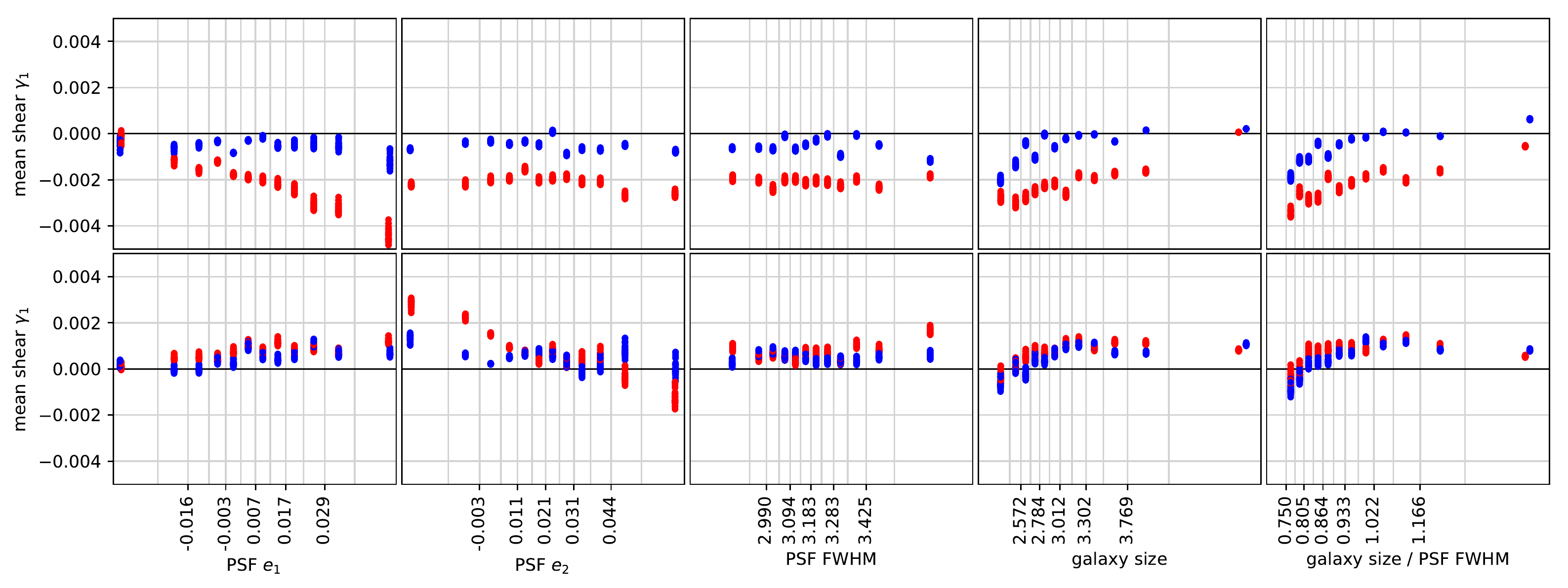}
\includegraphics[width=1\linewidth]{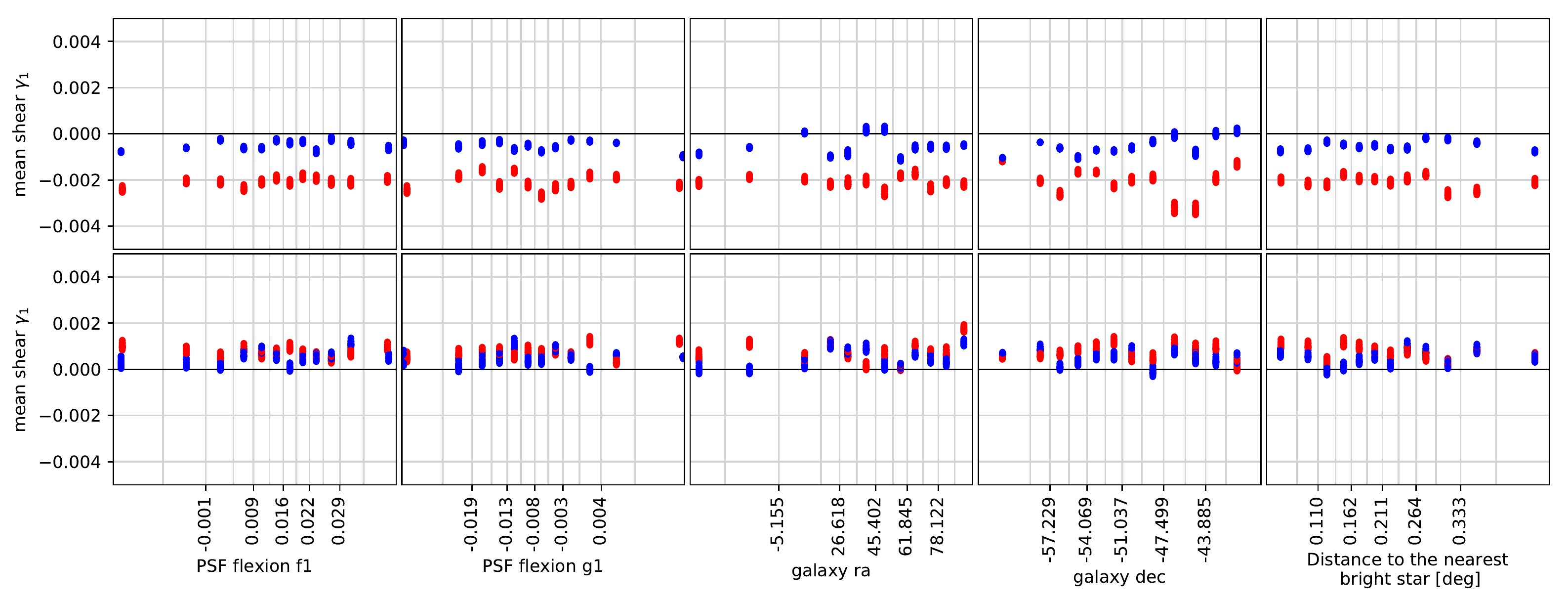}
\includegraphics[width=1\linewidth]{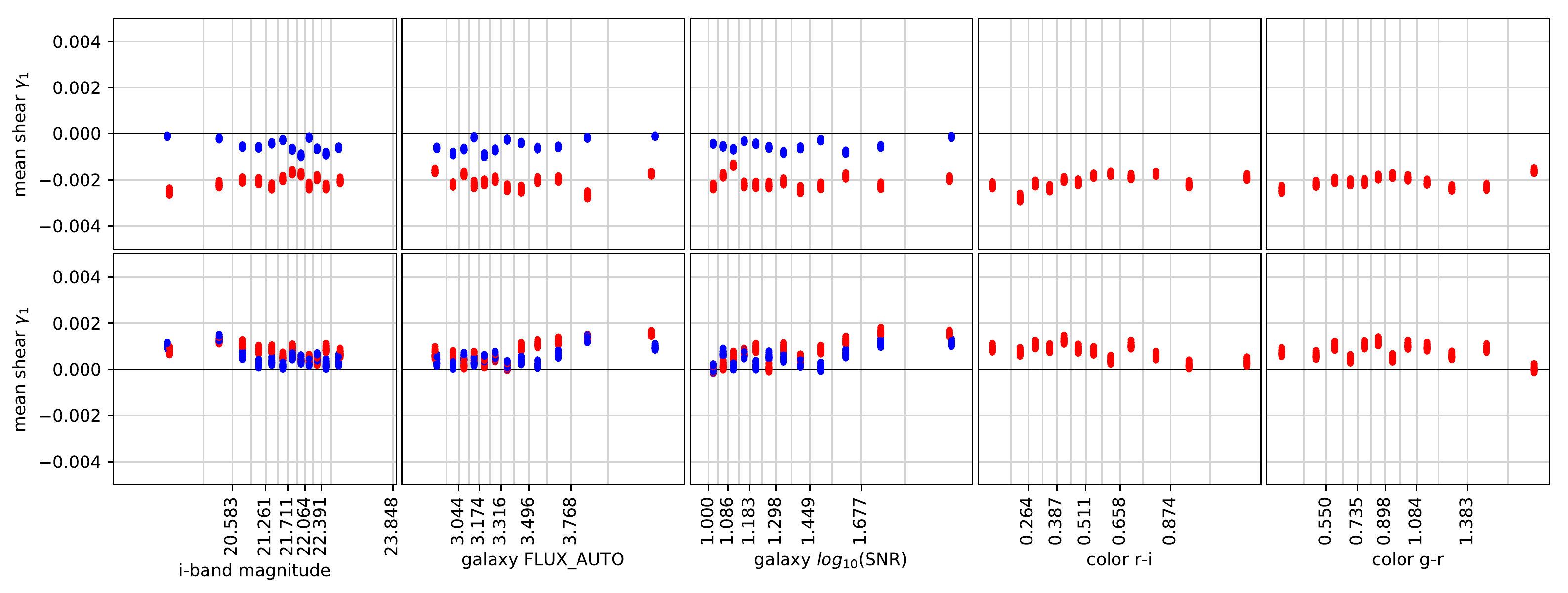}
\caption{Mean shear as a function of various parameters.
Galaxy size was calculated using \textsc{SExtractor} weighted moments with Equation~\ref{eqn:galaxy_size}.
}
\label{fig:psf_leakage}
\end{figure*}

We compare the DES and UFig catalogues in terms of the mean shear of the entire sample as a function of various properties.
The mean shear for the full sample is given in Section~\ref{sec:remaining_issues} and is much larger than expected from cosmic variance.
In our method, the mean shear is subtracted from the maps before power spectrum estimation.
This information would not be used anyway, as it contributes only to $\ell<100$.
Because of this, we can tolerate mean shear in the data, as long as the 2-pt statistics, and especially the B-mode, remain low.
Nevertheless, a statistically significant mean shear would be an indication of possible remaining issues.
In Figure~\ref{fig:psf_leakage} we investigate the behaviour of the mean shear as a function of various parameters: PSF shape, size, flexions, position in the survey footprint, brightness and signal-to-noise, colours, and distance to near bright star of magnitude $<$12, taken from the \textsc{Sky2000} catalogue \citep{Myers2001sky2000}.
We plot only $f_1$ and $g_1$ flexion, as the other components looked very similar.
Colours are plotted only for the DES data, as we simulate the full survey area only for the \band-band.
We plot only the $r-i$ and $g-r$ colours here, other combinations looked similar.
We also do not plot dependence on the PSF kurtosis and ration of the Moffat components, as the mean shear as a function of these parameters does not display any trends.

A trend that differs between the DES data and the UFig simulations would be an indication of a remaining issue that was not properly accounted for in the analysis or not modelled correctly in the simulations.
We notice a PSF leakage on the level of $\approx 4\%$ in the DES data.
We do not expect it to affect the measurement significantly, as described in Section~\ref{sec:systematics_model}.
Mean shear in the $\gamma_1$ direction is consistently low, and does not seem to depend in a different way than the simulations on any of the variables considered.
This suggests that the mean shear behaves purely like a constant offset, which would not have an influence on the shear power spectrum.

\section{Noise correlation}
\label{appendix:noise_correlation}

\begin{figure}[t!]
\includegraphics[width=0.5\linewidth]{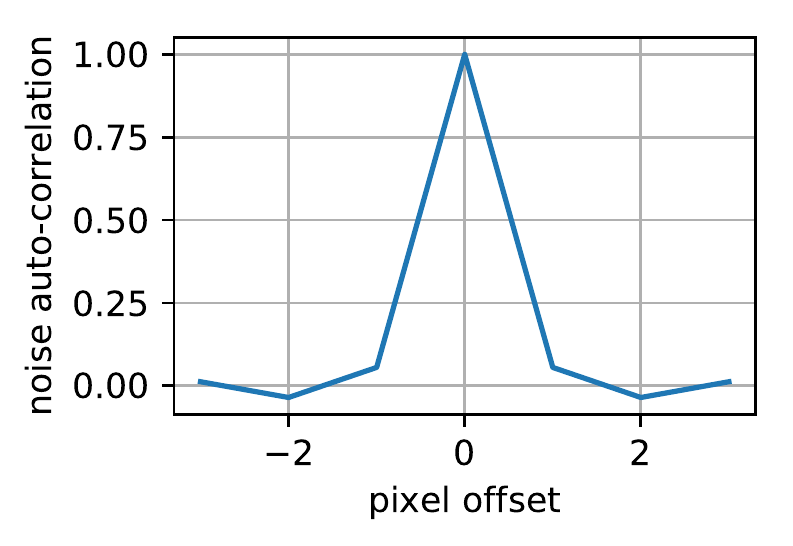}
\caption{Convolution kernel used to introduce noise correlation in the simulated images.}
\label{fig:noise_autocorr}
\end{figure}

The process of image resampling using the Lanczos kernel causes noise to become correlated.
We include this correlation in the \textsc{UFig} simulations by convoving the noise image with a specially designed kernel.
We create this kernel such that its auto-correlation is matched to that expected from Lanczos-resampled images.
We calculated this expected auto-correlation by measuring the average auto-correlation of a set of 20000 resampled noise images of size 100 $\times$ 100 pixels.
These images were resampled to a shifted coordinate system, with a random uniform shift of maximally $\pm 0.5$ pixel in both x and y directions.
The final kernel image calculated this way was contained in 7 $\times$ 7 pixel stamp, shown in Figure~\ref{fig:noise_autocorr}.

\bibliography{des_y1_shear_cosmology}

\end{document}